\documentclass[fleqn,10pt]{wlscirep}
\usepackage[utf8]{inputenc}
\usepackage[T1]{fontenc}
\usepackage{siunitx}
\usepackage{amsmath,amssymb}
\usepackage{graphicx}
\usepackage{gensymb}
\usepackage{csquotes} 

\title{Detection of ultra-high-energy cosmic rays in the southern hemisphere with FAST: data acquisition and preliminary results}

\author[1,2]{Jakub~Kmec}
\author[1,2]{Petr~Boril}
\author[3]{Fraser~Bradfield}
\author[1]{Karel~Cerny}
\author[2]{Ladislav~Chytka}
\author[3]{Toshihiro~Fujii}
\author[1]{Pavel~Horvath}
\author[1]{Miroslav~Hrabovsky}
\author[1,2]{Vlastimil~Jilek}
\author[1]{Jiri~Kvita}
\author[4]{Max~Malacari}
\author[5]{Massimo~Mastrodicasa}
\author[6]{John~N.~Matthews}
\author[1,2]{Stanislav~Michal}
\author[7]{Marcus~Niechciol}
\author[2]{Libor~Nozka}
\author[2]{Miroslav~Palatka}
\author[1,2]{Miroslav~Pech}
\author[4]{Paolo~Privitera}
\author[5,9]{Francesco~Salamida}
\author[3]{Shunsuke~Sakurai}
\author[2]{Petr~Schovanek}
\author[4]{Radomir~Smida}
\author[1,2]{Zuzana~Svozilikova}
\author[3]{Haruka~Tachibana}
\author[8]{Akimichi~Taketa}
\author[6]{Stan~B.~Thomas}
\author[1,2]{Petr~Travnicek}
\author[2]{Martin~Vacula}
\author[2]{Jiri~Zahora}
\author[1,2,b]{Dusan~Mandat}
\author[1,2,a]{Petr~Hamal}

\affil[1]{Palacký University Olomouc, Faculty of Science, Joint Laboratory of Optics of Palacký University and Institute of Physics of the Czech Academy of Sciences, 17. listopadu 1192/12, 779 00 Olomouc, Czech Republic}
\affil[2]{Institute of Physics of the Czech Academy of Sciences, Joint Laboratory of Optics of Palacký University and Institute of Physics of the Czech Academy of Sciences, 17. listopadu 1154/50a, 779 00 Olomouc, Czech Republic}
\affil[3]{Graduate School of Science, Osaka Metropolitan University, Sumiyoshi-ku, Osaka, Japan}
\affil[4]{Kavli Institute for Cosmological Physics, University of Chicago, Chicago, IL, USA}
\affil[5]{Department of Physical and Chemical Science, University of L’Aquila, L'Aquila, Italy}
\affil[6]{High Energy Astrophysics Institute and Department of Physics and Astronomy, University of Utah, Salt Lake City, UT, USA}
\affil[7]{Center for Particle Physics Siegen, University of Siegen, Germany}
\affil[8]{Earthquake Research Institute, University of Tokyo, Bunkyo-ku, Tokyo, Japan}
\affil[9]{INFN Laboratori Nazionali del Gran Sasso, L'Aquila, Italy}
\affil[a]{p.hamal@upol.cz}
\affil[b]{mandat@fzu.cz}

\begin{abstract}
Ultra-high-energy cosmic rays (UHECRs) remain one of the greatest mysteries in astroparticle physics.
The Fluorescence detector Array of Single-pixel Telescopes (FAST) is a next-generation cosmic ray experiment which utilizes ground-based fluorescence telescopes designed to detect these extremely rare particles at energies exceeding $30\, \unit{EeV}$.
FAST offers a cost-effective and low-maintenance solution to cover the huge detection areas required for UHECR observation.
FAST telescopes are currently installed and remotely operated in both hemispheres, at the Pierre Auger Observatory and the Telescope Array experiment. 
To enable fully autonomous operation, a sophisticated trigger for data acquisition is essential.
In this paper, we present two novel triggering algorithms inspired by those used at the largest observatories, but improved to meet the specific requirements imposed by the FAST design.
Their performance is validated using Monte Carlo simulations of extensive air showers and UHECR events detected by the FAST telescope in the southern hemisphere.
Finally, we present the sensitivity analysis estimate for FAST. 
\end{abstract}

\begin{document}
\flushbottom
\maketitle
\thispagestyle{empty}

\section{Introduction}
Ultra-high-energy cosmic rays (UHECRs) -- defined in this paper as cosmic ray particles with energies exceeding $10^{17} \,\unit{eV}$ ($0.1\,\unit{EeV}$) -- are believed to originate from the most energetic phenomena in the Universe \cite{Hillas1984, Coleman2023}. 
When a high-energy cosmic ray interacts with an atmospheric nucleus, it produces a cascade of secondary particles known as an extensive air shower (EAS).
By studying the properties of the EAS, it is possible to determine the energy, arrival direction, and mass-sensitive information of the corresponding UHECR. 
Cosmic rays with energies greater than $100\,\unit{EeV}$ are extremely rare, arriving at Earth with a rate of approximately one particle per square kilometer per century.
As a result, modern observatories are designed to cover huge detection areas and operate for many years to collect a large number of events.

Two well-established detection techniques are commonly used~\cite{Coleman2023}: 
(1) detection via ground-based particle detectors \cite{TA2012}, which involves measuring EAS particles at ground level using instruments such as water-Cherenkov stations or plastic scintillators, 
and (2) detection of air fluorescence using fluorescence telescopes \cite{Auger2010}, which capture UV fluorescence light in the $290-430\,\unit{nm}$ wavelength range emitted by atmospheric nitrogen excited by EAS particles \cite{Ave2008, Keilhauer2013}.
The two largest observatories, the Pierre Auger Observatory (Auger) \cite{Auger2015} and the Telescope Array experiment (TA) \cite{TA2008}, employ a hybrid approach that combines both techniques. 
Auger, located near the city of Malarg\"ue in Argentina, covers approximately $3000\,\unit{km}^2$, while TA, situated near the city of Delta in Utah, USA, covers around $700\,\unit{km}^2$. 
Other techniques for detecting high-energy cosmic rays include, for example, measuring the radio emission from EASs \cite{Aab2018, Alvarez2019} or detecting individual tracks of EAS particles in charged couple devices \cite{Kawanomoto2023}.
Despite the significant progress in understanding cosmic rays over recent decades, thanks to Auger, TA, and other experiments such as the IceCube Neutrino Observatory in Antarctica, which is dedicated to detecting Cherenkov light \cite{Aartsen2017}, the origin and nature of UHECRs remain unknown.
This makes UHECRs one of the greatest mysteries in astroparticle physics~\cite{Watson2014, Coleman2023}.

The most energetic cosmic rays ever detected include a $320\,\unit{EeV}$ particle observed in 1991 \cite{Bird1995}, a $280\,\unit{EeV}$ particle detected in 2001 \cite{Chikawa2001}, and the most recent $244\,\unit{EeV}$ particle observed in 2021 \cite{TA2023}.
The first particle was detected with a fluorescence telescope, while the latter two were detected using plastic scintillators.
Since extragalactic cosmic ray particles interact with the cosmic microwave background radiation, they cause energy losses during their propagation through space.
The energy losses lead to a strong suppression of the UHECR flux above the so-called Greisen-Zatsepin-Kuzmin (GZK) cutoff \cite{Greisen1966, Zatsepin1966}, and limit the maximum distance from which UHECRs can reach Earth without suffering significant attenuation, the so-called GZK horizon.
For protons, the GZK cutoff occurs at approximately $50\,\unit{EeV}$, and the corresponding GZK horizon is typically in the range of $50-100\,\unit{Mpc}$ \cite{TA2023, Coleman2023}.

To overcome the challenge posed by the extremely low flux of particles with the highest energies, it is necessary to construct an observatory 
covering huge areas to give the exposure required for measuring a very large number of UHECRs above $100\,\unit{EeV}$.
The Fluorescence detector Array of Single-pixel Telescopes (FAST) is an experiment which has been designed to meet this requirement.
It will make use of a cost-effective and low-maintenance ground-based fluorescence detector array, composed of multiple stations of FAST telescopes (12 telescopes per station) arranged in a triangular grid \cite{Fujii2016, Mandat2017, Malacari2020}.
For a detailed description of the optical, mechanical and electrical design, we refer the reader to \cite{Mandat2017}.
In the following, we provide only a brief overview.

\begin{figure}[htb!]
    \centering
    \includegraphics[width=0.28\linewidth]{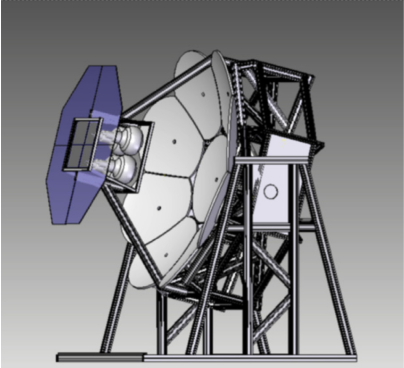}
    \caption{Schematic of the first-generation FAST telescope \cite{Fujii2016}.}
    \label{fig:schematic}
\end{figure}

Each FAST telescope consists of four photomultiplier tubes (PMTs), with each PMT covering an angular field of view of approximately $\Delta \Omega = 15\degree$.
The PMTs are positioned at the focal plane of a segmented spherical mirror with a diameter of $1.6\,\unit{m}$,
providing a total field of view of $30\degree \times 30\degree$ in azimuth and elevation with an effective light-collecting area of $1\,\unit{m}^2$.
Despite its simplified design, FAST has the potential to achieve resolutions in energy and $X_{\mathrm{max}}$ (the depth of the EAS maximum) comparable to those of the Auger and TA observatories at energies above $30\, \unit{EeV}$ \cite{Fujii2016, Bradfield2025}. 
A~schematic of the first-generation FAST telescope is shown in Fig. \ref{fig:schematic}.

Currently, five first-generation FAST telescopes are installed, four of which are operated remotely:
three at TA \cite{Malacari2020} and one at Auger \cite{Hamal2024, Bradfield2024}. 
Both TA/Auger provide power and wireless internet connectivity to the FAST telescopes, as well as external triggers from adjacent telescopes with overlapping fields of view.
A second-generation FAST telescope has recently been developed, featuring an improved segmented mirror and a new enclosure equipped with a solar power system.
With a new data acquisition system using an independent internal trigger, the new FAST telescopes are capable of fully autonomous operation \cite{Hamal2024}. 
By 2026, the FAST collaboration aims to deploy a mini-array of six FAST telescopes at Auger, arranged in a triangular configuration \cite{Hamal2024}. 
This setup will enable full reconstruction of event geometry, energy, and $X_{\mathrm{max}}$ through stereo observation. 
Since identical fluorescence telescopes will be used in both hemispheres, one of FAST's primary future goals is to help resolve the energy difference observed between Auger and TA at the highest energies \cite{Plotko2023}.
Furthermore, FAST is being considered as a cost-effective fluorescence experiment for the proposed Global Cosmic Ray Observatory (GCOS) \cite{Horandel2022, Coleman2023}, which aims to cover an area exceeding $60,\!000\,\unit{km}^2$ \cite{Ahlers2025}.

Given that FAST is designed as a low-cost observatory, it differs in two major ways from the largest existing observatories.
First, the low resolution of the FAST telescope, combined with its relatively small effective light-collecting area ($A = 1\,\unit{m^2}$), means that it can only reliably detect events with higher energies. 
The signal-to-noise ratio (SNR) is proportional to $\sqrt{A/\Delta \Omega}$, where $\Delta \Omega$ is the pixel solid angle, i.e., the field of view of a single PMT.
Compared to the fluorescence detectors of Auger ($A = 3.00\,\unit{m^2}$, $\Delta \Omega = 1.5\degree$) and TA ($A = 5.95\,\unit{m^2}$, $\Delta \Omega = 1.0\degree$), the SNR from a given shower is significantly lower for FAST.
Specifically, the SNR is approximately $9.45$ times lower than TA and $5.48$ times lower than Auger.
In this regard, FAST is more comparable to the Fly's Eye experiment ($A = 1.70\,\unit{m^2}$, $\Delta \Omega = 5.5\degree$), which operated between 1981 and 1993 \cite{Bird1994}.
The second key difference lies in the event reconstruction process.
Traditional fluorescence detection techniques use a bottom-up approach, where the timing information from a track of triggered pixels is used to fit the shower geometry \cite{Auger2010}. 
However, this method is not feasible for FAST, as each telescope consists of only four PMTs compared to the several hundred PMTs in conventional fluorescence telescopes.
Instead, FAST uses a top-down approach, where the shower parameters are determined by comparing the time series from each PMT (including those without a significant signal from an EAS) with those from simulated events using a maximum likelihood estimation~\cite{Malacari2020, Bradfield2024}.

A sophisticated triggering system is essential for effective data acquisition. 
Auger and TA implement similar internal trigger algorithms composed of multiple hardware levels \cite{TA2009, Auger2010}.
The hardware triggering system operates in two main stages.
The first level identifies potential signals at individual PMTs by applying moving average or moving sum filters, the results of which are then compared to thresholds based on the standard deviation of the background noise.
A PMT triggers if the averaged or summed value exceeds the predefined threshold.
While both experiments use this approach, there are key differences in implementation.
TA applies moving averages over four time windows ($1.6, 3.2, 6.4, 12.8\,\mu\unit{s}$) \cite{TA2009}, whereas Auger uses a moving sum with a single fixed time window selected within the range $0.5 - 1.6\,\mu\unit{s}$ \cite{Auger2010, Albury2020}.
Additionally, TA employs a fixed threshold, resulting in an averaged pixel trigger rate of approximately $3\,\unit{Hz}$, while Auger dynamically adjusts its threshold to maintain a constant pixel trigger rate of $100\,\unit{Hz}$.
The second level processes the information from the triggered pixels from the first level. 
Although the exact implementations again differ between TA and Auger, the fundamental idea is to search for spatial patterns of triggered pixels.
This is performed by identifying adjacent triggered PMTs that match predefined patterns, with the required number of triggered pixels ranging from $3$ to $5$, depending on the PMT positions and algorithm specifications \cite{TA2009, Auger2010}.
Events that pass the hardware trigger can be further filtered using software to exclude non-EAS events such as lightning, muons, airplanes, or randomly triggered pixels.
A comparable data acquisition strategy was used in the Fly’s Eye experiment \cite{Bird1994}, which employed three time windows to maintain a constant $50\,\unit{Hz}$ trigger rate for each PMT.
For an individual telescope (referred to as a mirror) to be triggered, at least two PMTs were required to trigger simultaneously.

For fully autonomous operation with FAST, a high-quality internal trigger must be developed.
While inspiration can be drawn from the first-level trigger algorithms used in Auger and TA, the significantly lower SNR in FAST requires enhancements to ensure robustness of the trigger.
This is especially important because the FAST telescope, with only four PMTs, is unable to use the pattern of triggered pixels as a method for removing potential false triggers.
This presents a significant challenge for the design of the trigger system implementation in hardware.
In this paper, we introduce a novel triggering approach inspired by the first-level trigger of Auger and TA,
adapted to meet the specific requirements of individual FAST telescopes.
The proposed method is optimized for signal detection at standalone FAST telescopes and designed to support data acquisition for the future FAST mini-array at the Auger observatory.
We evaluate its performance through Monte Carlo simulations of EAS and compare it with existing approaches using both simulated data and UHECR events detected by the FAST telescope in the southern hemisphere via the external trigger.
Finally, we present the sensitivity analysis estimate for FAST.

\section{Methods}
In this section, we first motivate the need for new approaches to data acquisition in Sect.~\ref{subsec:motivation} followed by an introduction to the trigger algorithms in Sect.~\ref{subsec:algorithms}.
Section~\ref{subsec:baselines} investigates the characteristics of the noise background, which is then used to calculate trigger thresholds in Sect.~\ref{subsec:thresholds}.

\subsection{Motivation}
\label{subsec:motivation}
The time series recorded by each PMT are defined in units of ADC counts and are referred to as traces. 
For clarity, the term signal refers to the signal of an EAS throughout this manuscript.
The triggering algorithm will be applied at the hardware level using ADC counts; however, for development and testing in this manuscript, traces are scaled to the estimated number of photoelectrons $N_{\mathrm{p.e.}}$ to allow direct comparison across PMTs.
This scaling was previously performed by calibration using a light source of known flux in dedicated laboratory measurements.
Although an additional gain calibration is planned to account for PMT aging, this will not affect the results presented here, either for the comparison of triggering algorithms or for the UHECRs detected by the FAST telescope.

\begin{figure}[htb!]
    \centering
    \includegraphics[width=0.80\linewidth]{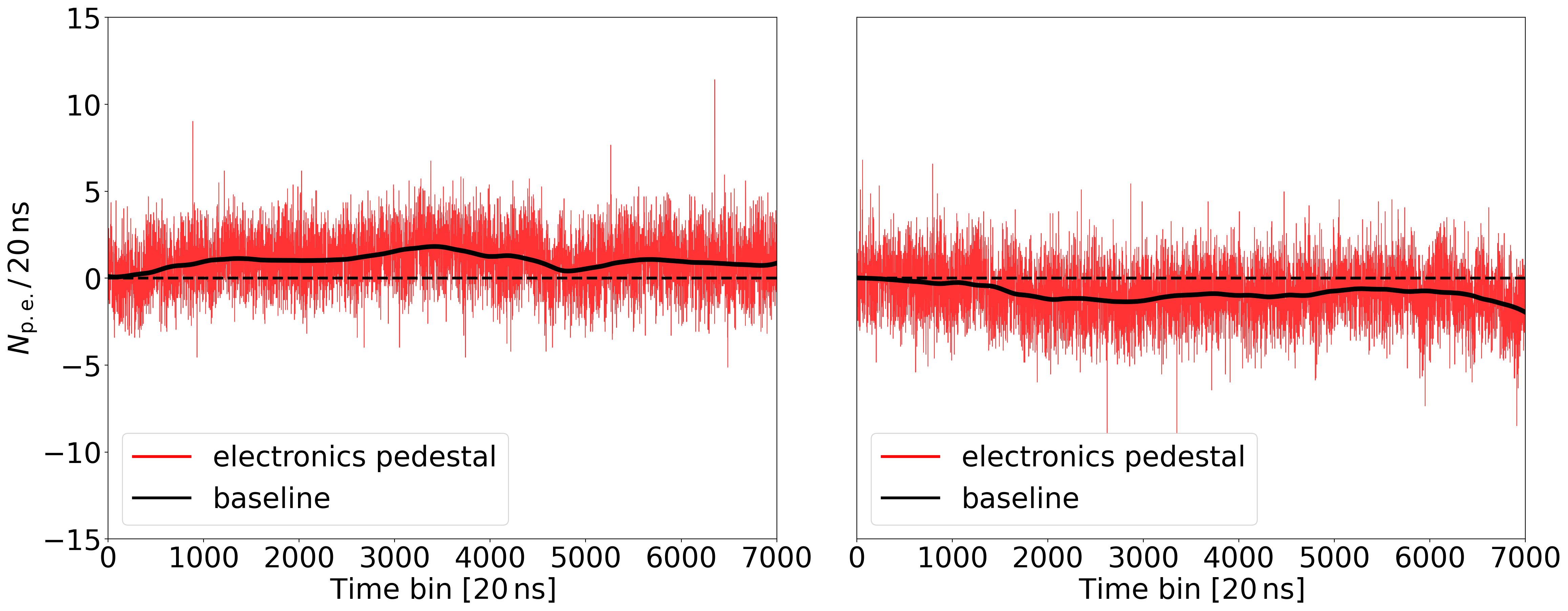}
    \caption{Examples of acquired floating pedestals and their extracted baselines, recorded with the telescope shutter closed.
    The $x-$axis represents time, and the $y-$axis shows the estimated number of photoelectrons per $20 \, \unit{ns}$.
    The baseline can either overestimate (left panel) or underestimate (right panel) a potential signal.}
    \label{fig:motivation_01}
\end{figure}

A major challenge in data acquisition is the slowly varying nature of the electronics noise, which is referred to as the electronics pedestal.
This effect is illustrated in Fig. \ref{fig:motivation_01}, which shows two examples of pedestal behavior.
The pedestals were recorded with the PMTs under high voltage and the telescope shutter closed, eliminating potential signals from EAS and any contribution from the night sky background -- thus, only the electronics noise was recorded.
Each pedestal trace represents one PMT and consists of $7000$ bins, with each bin corresponding to $20 \, \unit{ns}$.
The average value of the first $500$ bins was subtracted to shift the pedestal to zero.
The black lines in Fig. \ref{fig:motivation_01} represent the baselines and their extraction is described in Sect.~\ref{subsec:baselines}.
In an ideal case, the baseline remains constant, but for FAST it floats, as clearly seen in Fig. \ref{fig:motivation_01}.
Therefore, the terms ``floating baseline'' or ``floating pedestal'' are often used in the text to emphasize this behavior.
Two distinct cases can be observed: when the baseline is above zero (left panel of Fig.~\ref{fig:motivation_01}), the signal of a potential event would be overestimated; when the baseline is below zero (right panel of Fig.~\ref{fig:motivation_01}), the signal would be underestimated.

\begin{figure}[htb!]
    \centering
    \includegraphics[width=0.80\linewidth]{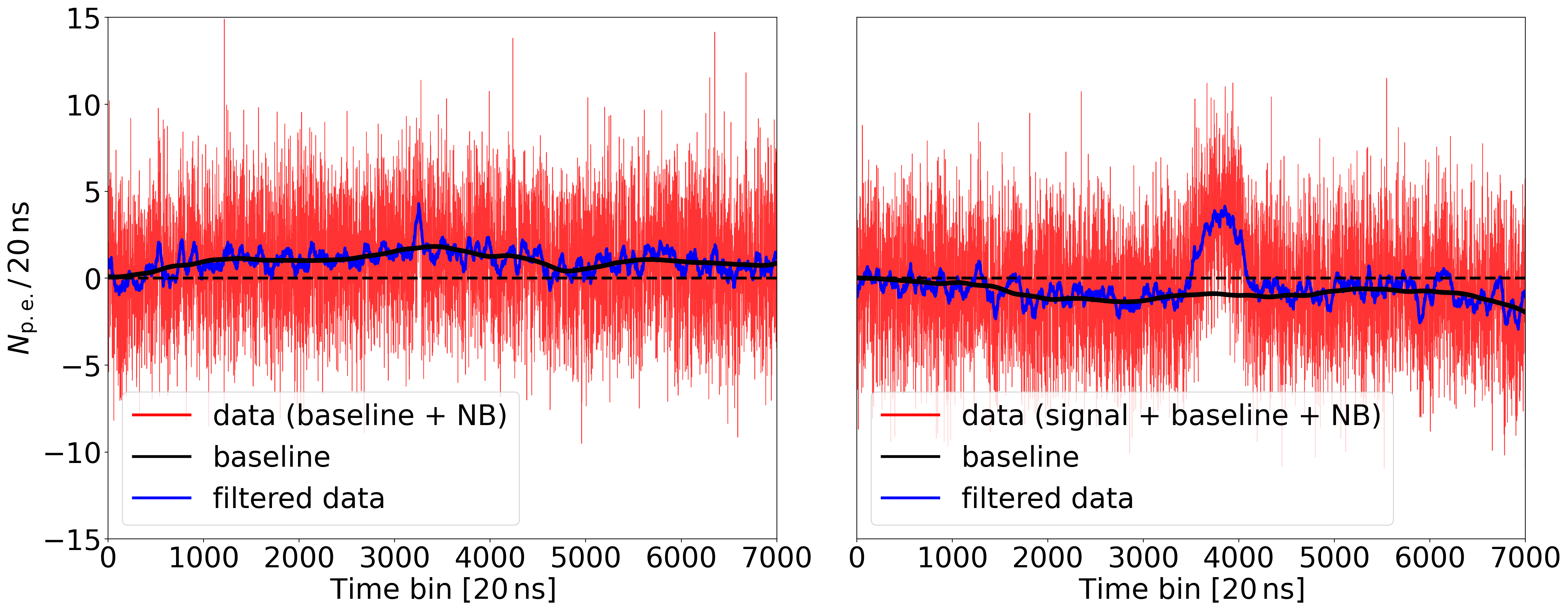}
    \caption{Left panel: The baseline combined with a simulated NB, i.e., it does not contain a signal. Right panel: The baseline combined with a simulated NB and a signal.
    The filtering process based on the Auger first-level trigger method is applied for both panels.
    Due to the floating baseline, the filtered data without a signal in the left panel reaches a higher maximum value ($\approx 4.30\,N_{\mathrm{p.e.}}/20\,\unit{ns}$) than in the case when a signal is present in the right panel ($\approx 4.14\,N_{\mathrm{p.e.}}/20\,\unit{ns}$).}
    \label{fig:motivation_02}
\end{figure}

This baseline variability can therefore cause noise to appear as a signal or, conversely, cause the signal of an EAS to be missed.
The effect is demonstrated in Fig. \ref{fig:motivation_02}, where both baselines from Fig. \ref{fig:motivation_01} are combined with a simulated noise background (NB).
The NB consists of contributions from the night sky background and electronics noise,
and during regular data acquisition, it typically behaves as Gaussian noise. 
Therefore, we model it as $\mathrm{NB} \sim \mathcal{N}(\mu_{\mathrm{NB}},\,\sigma_{\mathrm{NB}}^{2})$ with $\mu_{\mathrm{NB}} = 0$ and $\sigma_{\mathrm{NB}} = 2.5\,N_{\mathrm{p.e.}}/20\,\unit{ns}$, which reflects typical night conditions under a clear sky.
Here, $\mathcal{N}$ denotes the Gaussian distribution.
The combination of the simulated NB with the extracted floating baseline provides a realistic representation of a real NB affected by the floating pedestal during FAST data acquisition.

In the left panel of Fig. \ref{fig:motivation_02}, the baseline from the left panel of Fig. \ref{fig:motivation_01} is combined with a simulated NB, representing a ``no-signal trace''.
The right panel of Fig. \ref{fig:motivation_02} shows the baseline from the right panel of Fig. \ref{fig:motivation_01} combined with a simulated NB and an additional simulated signal of an EAS, representing a ``signal trace''.
For the ``no-signal trace'', the NB was generated to include a minor peak near the center. 
To evaluate whether a signal is present in Fig. \ref{fig:motivation_02}, the simulated data (shown in red) are filtered using a moving average (MA) with a window length of $51$ bins, producing the filtered traces shown in blue.
This method reproduces the approach of the first-level trigger in the Auger experiment,
with the window length matching that in Auger \cite{Albury2020}.
The filtered ``no-signal trace'' reaches a higher maximum value ($\approx 4.30\,N_{\mathrm{p.e.}}/20\,\unit{ns}$) than the filtered ``signal trace'' ($\approx 4.14\,N_{\mathrm{p.e.}}/20\,\unit{ns}$).
This points to a fundamental issue for FAST: a floating baseline can cause a noise-only trace to appear stronger than one containing a signal when the previous method is applied. 
Consequently, real signals may drop below the detection threshold due to the presence of floating baselines and therefore be missed.

\subsection{Algorithm definitions}
\label{subsec:algorithms}
The challenge described above arises from the significantly lower SNR in FAST compared to the TA and Auger experiments. 
To address this, it is necessary to account for the presence of floating baselines.
In this section, we present four different triggering algorithms.
Two of them are developed by the authors and are referred to as \textit{in-house} algorithms.
The other two, referred to as \textit{reference} algorithms, are similar to those used by the Auger and TA experiments and are included for comparison.

The basic principle of SNR calculation involves applying a filter to the trace (e.g., an MA filter) to obtain its smoothed version.
The SNR is then typically computed as a ratio of the filtered trace and its standard deviation.
However, when implementing such a calculation in hardware, only simple mathematical operations such as addition, subtraction, and multiplication can be used.
To meet these constraints, a low-pass Finite Impulse Response (FIR) filter is employed to produce the filtered trace.
The FIR filter is defined by the discrete convolution of the input trace $x_{n-k}$ with the impulse response $h_k$ \cite{Jackson2013}:
\begin{equation}
    y_n = \sum_{k = 0}^{m-1} h_k \cdot x_{n-k},
    \label{eq:FIR}
\end{equation}
where $y_n$ is the output at position $n$, $x_{n-k}$ is the input at position $n-k$, $h_k, k = 0, \ldots, m - 1$ are the FIR filter coefficients, and $m$ is the filter length. 
In particular, the MA filter is obtained by setting $h_k = \frac{1}{m}$ for all $k$.

A general SNR formula for our in-house algorithms is written as follows:
\begin{equation}
    \mathrm{SNR} = \frac{  \mathrm{trace}_{\mathrm{filt}} (P_{\mathrm{test}})  - \mathrm{trace}_{\mathrm{mean}}(P_{\mathrm{ma}}) }
    {\sigma(\mathrm{trace}_{\mathrm{filt}})},
    \label{eq:snr}
\end{equation}
where $\mathrm{trace}_{\mathrm{filt}} (P_{\mathrm{test}})$ is the value of the FIR-filtered trace given by Eq. (\ref{eq:FIR}) at the tested position $P_{\mathrm{test}}$, $\mathrm{trace}_{\mathrm{mean}}(P_{\mathrm{ma}})$ is the estimated baseline value at the position $P_{\mathrm{ma}}$, and $\sigma(\mathrm{trace}_{\mathrm{filt}})$ is the standard deviation of the filtered trace.
The details of the baseline and standard deviation calculations are given below.

\begin{figure}[htb!]
    \centering
    \includegraphics[width=0.43\linewidth]{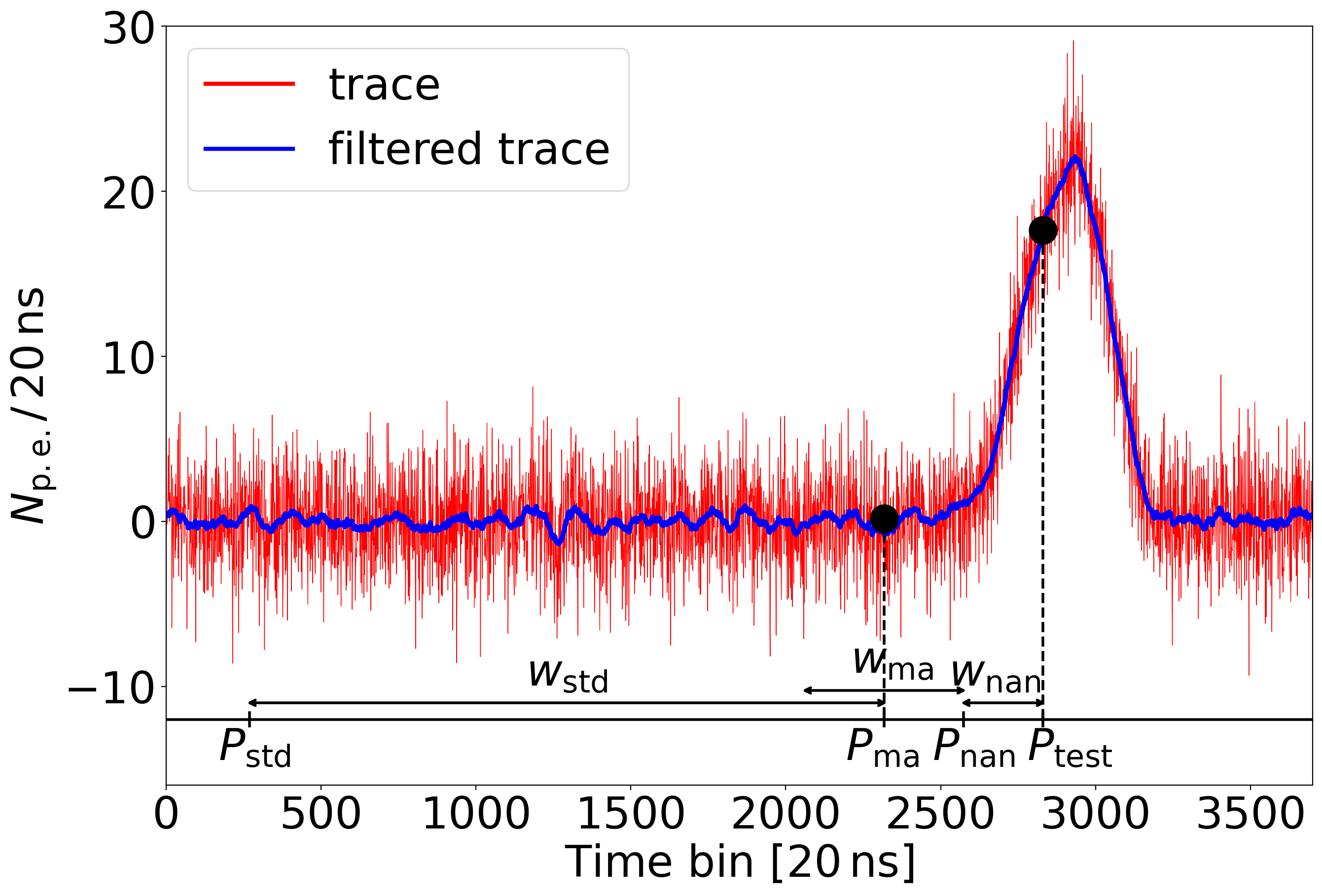}
    \caption{A general scheme for SNR calculation. $P_{\mathrm{test}}$ denotes the current tested position, and $P_{\mathrm{ma}}$ is the position used to estimate the baseline. The standard deviation of the filtered trace is computed over the interval between $P_{\mathrm{std}}$ and $P_{\mathrm{ma}}$. Values between $P_{\mathrm{nan}}$ and $P_{\mathrm{test}}$ are excluded from both the baseline and standard deviation calculations.
    In this example, the parameters are $w_{\mathrm{std}} = 2048$, $w_{\mathrm{ma}} = 513$, and $w_{\mathrm{nan}} = 256$ bins.}
    \label{fig:alg_scheme}
\end{figure}

The SNR calculation scheme is illustrated in Fig. \ref{fig:alg_scheme}.
The raw trace is shown in red, and the filtered trace in blue.
The FIR-filtered value at $P_{\mathrm{test}}$ and the baseline estimate $\mathrm{trace}_{\mathrm{mean}}$ at $P_{\mathrm{ma}}$ are indicated with black dots.
The standard deviation is calculated over the interval $[P_{\mathrm{std}}, P_{\mathrm{ma}}]$, which has length $w_{\mathrm{std}}$ bins.
The baseline estimate $\mathrm{trace}_{\mathrm{mean}}$ at $P_{\mathrm{ma}}$ 
is computed by averaging trace values in the interval $[P_{\mathrm{nan}} - w_{\mathrm{ma}}, P_{\mathrm{nan}}]$, where $w_{\mathrm{ma}}$ is the number of bins averaged.
Values in the interval $[P_{\mathrm{nan}}, P_{\mathrm{test}}]$, of total length $w_{\mathrm{nan}}$ bins, are excluded from both the baseline and standard deviation computations to avoid potential signal contamination.
The window lengths $w_{\mathrm{std}}, w_{\mathrm{ma}}, w_{\mathrm{nan}}$ are parameters of the SNR model.
For brevity, units ``bins'' will be omitted in the following text.

Due to the linearity of the FIR filter, the standard deviation of the filtered trace $\sigma(\mathrm{trace}_{\mathrm{filt}})$ in Eq.\,(\ref{eq:snr}) can be calculated directly from the unfiltered trace using: 
\begin{equation}
    \sigma(\mathrm{trace}_{\mathrm{filt}}) = \sigma(\mathrm{trace})\sqrt{\sum_{k = 0}^{m-1} h_k^2},
    \label{eq:sigma}
\end{equation}
where $\sigma(\mathrm{trace})$ is the standard deviation of the unfiltered trace.
For derivation, see Appendix \ref{appendix:calib00_mean}.
Using Eq.\,(\ref{eq:sigma}) in Eq.\,(\ref{eq:snr}) offers two major advantages. 
First, it simplifies algorithm’s hardware implementation, as the standard deviation can be calculated from the original trace instead of the filtered one.
Second -- and more importantly -- $\sigma(\mathrm{trace})\sqrt{\sum_{k = 0}^{m-1} h_k^2}$ exhibits substantially lower variability than $\sigma(\mathrm{trace_{\mathrm{filt}}})$, leading to a more reliable estimation of the standard deviation.
This approach is employed in the SNR calculation for Auger (see Eq.\,(3.25) in \cite{Albury2020}); however, it is not clear for TA, as this is not specified in their triggering paper \cite{TA2009}.
Moreover, we were unable to find a rigorous mathematical proof that $\sigma(\mathrm{trace_{\mathrm{filt}}})$ indeed has greater variability.
We consider this characteristic important; thus, a numerical analysis of this behavior is presented in Appendix \ref{appendix:std}.

In order to identify the most effective triggering method for FAST, we now define the four SNR algorithms in Eqs.\,(\ref{eq:snr1})-(\ref{eq:snr4}), all of which are derived from Eq.\,(\ref{eq:snr}):
\begin{subequations} \label{eq:snr_definitions}
\begin{align}
    & \mathrm{SNR}^{(1)}_{\mathrm{inhouse}} = \frac{  \mathrm{trace}_{\mathrm{FIR}} (P_{\mathrm{test}})  - \mathrm{trace}_{\mathrm{mean}}(P_{\mathrm{ma}}) }
    {\sigma(\mathrm{trace})\sqrt{\sum_{k = 0}^{m-1} h_k^2}}, \label{eq:snr1} \\
    & \mathrm{SNR}^{(2)}_{\mathrm{inhouse}} = \frac{  \mathrm{trace}_{\mathrm{MA}} (P_{\mathrm{test}})  - \mathrm{trace}_{\mathrm{mean}}(P_{\mathrm{ma}}) }
    {\sigma(\mathrm{trace})/\sqrt{m}}, \label{eq:snr2} \\
    & \mathrm{SNR}^{(1)}_{\mathrm{reference}} = \frac{  \mathrm{trace}_{\mathrm{MA}} (P_{\mathrm{test}}) }
    {\sigma(\mathrm{trace})/\sqrt{m}}, \label{eq:snr3} \\
    & \mathrm{SNR}^{(2)}_{\mathrm{reference}} = \frac{  \mathrm{trace}_{\mathrm{MA}} (P_{\mathrm{test}}) }
    {\sigma(\mathrm{trace_{\mathrm{filt}}})}. \label{eq:snr4}
\end{align}
\end{subequations}

\noindent
The first algorithm inhouse$^{(1)}$, given in Eq.\,(\ref{eq:snr1}), uses a general FIR filter with coefficients~$h_k$.
The second algorithm inhouse$^{(2)}$, defined by Eq.\,(\ref{eq:snr2}), uses a specific case of the FIR filter, the MA filter.
The remaining two algorithms applied for FAST are based on the Auger/TA approaches and are denoted reference$^{(1)}$ and reference$^{(2)}$. 
They are obtained from Eq.\,(\ref{eq:snr}) by setting $\mathrm{trace}_{\mathrm{mean}}(P_{\mathrm{ma}}) = 0$, consequently ignoring the floating baseline.
These two variants use the MA filter, and they differ in how the standard deviation is computed: 
in Eq.\,(\ref{eq:snr3}), it is calculated from the raw trace, while in Eq.\,(\ref{eq:snr4}) it is computed from the filtered trace.
Eq.\,(\ref{eq:snr4}) is included to demonstrate that it is not appropriate to use $\sigma(\mathrm{trace_{\mathrm{filt}})}$.
Note that in the Auger first-level trigger \cite{Auger2010}, the standard deviation is not explicitly calculated;
instead, its effect is incorporated into a threshold.
This is possible through the use of a dynamic threshold in their implementation and is thoroughly discussed in Sect.~\ref{subsec:disc:std} in the Discussion section.
A comparison of the four algorithms is provided in Table \ref{tab:snr_definitions}.

\begin{table}[h!]
    \centering
    \caption{Comparison of algorithms.}
    \begin{tabular}{l||c|c|c}
    algorithm & filter & baseline estimation & standard deviation  \\
    \hline
    \hline
    inhouse$^{(1)}$  & general FIR    & yes & raw trace      \\
    inhouse$^{(2)}$  & MA & yes & raw trace      \\
    reference$^{(1)}$ & MA & no  & raw trace      \\
    reference$^{(2)}$ & MA & no  & filtered trace \\
    \end{tabular}
    \label{tab:snr_definitions}
\end{table}

During data acquisition, it is beneficial to apply multiple filter lengths to the trace, regardless of whether a general FIR filter or an MA filter is used \cite{TA2009}.
This approach is practical because signals of different durations require different filtering strategies.
Longer signals benefit from longer filter lengths, which more effectively suppress the NB.
Conversely, shorter signals require shorter filters, as longer ones may suppress or even eliminate the signal entirely.
In our implementation, we use five filter lengths $m$ (denoted $w_{\mathrm{filt}}$), each associated with its own variable threshold designed to maintain a trigger rate of $1.25\,\unit{Hz}$.
As a result, each PMT can reach a maximum trigger rate of $6.25\,\unit{Hz}$, leading to a maximum of $25\,\unit{Hz}$ for the entire FAST telescope, as it consists of four PMTs.
This represents an upper bound, since in practice a signal may trigger multiple PMTs and multiple filter lengths simultaneously, resulting in a slightly lower overall trigger rate.
The selected trigger rate is high enough to ensure sufficient sensitivity of FAST, but not so high as to produce an excessive data volume.

\begin{table}[h!]
    \centering
    \caption{Algorithm parameters given as number of bins (one bin corresponds to $20\,\unit{ns}$). The coefficients $h_k$ of the inhouse$^{(1)}$ algorithm are computed using a Hamming window.}
    \begin{tabular}{l||c|c|c|c|c}
    algorithm        & $w_{\mathrm{nan}}$ & $w_{\mathrm{ma}}$ & $w_{\mathrm{std}}$ & $w_{\mathrm{filt}}$ & cutoff frequency \\
    \hline
    \hline
    inhouse$^{(1)}$  & $256$  & $513$  & $2048$ & $25, 51, 101, 201, 401$ & $100\,\unit{kHz}$ \\
    inhouse$^{(2)}$  & $256$  & $513$  & $2048$ & $25, 51, 101, 201, 401$ &  -- \\
    reference$^{(1)}$ & $512$  & $0$    & $2048$ & $25, 51, 101, 201, 401$ &  -- \\
    reference$^{(2)}$ & $512$  & $0$    & $2048$ & $25, 51, 101, 201, 401$ &  -- \\
    \end{tabular}
    \label{tab:params}
\end{table}

The specific parameters used in all four algorithms are summarized in Table \ref{tab:params}.
For the reference algorithms, the baseline estimation is disabled by setting $w_{\mathrm{ma}} = 0$.
However, $w_{\mathrm{nan}} = 512$ is used to match the interval over which the standard deviation is computed in the in-house algorithms.  
For the FIR filter used in the inhouse$^{(1)}$ algorithm, the coefficients $h_k$ are computed using a Hamming window \cite{Jackson2013}, with a cutoff frequency of $100\,\unit{kHz}$ applied for all filter lengths $w_{\mathrm{filt}}$.
A detailed derivation of all parameter values, along with the frequency responses of the filters, is provided in Appendix \ref{appendix:calib}.

\subsection{Baseline generation}
\label{subsec:baselines}
To compare the different trigger algorithms defined in Eq.\,(\ref{eq:snr_definitions}), it is necessary to characterize the behavior of floating baselines.
To this end, pedestal data were acquired with PMTs under high voltage over the period from 11.8.2024 to 13.8.2024 from the single FAST telescope operational at Auger.
Since the contribution of the night sky background to the floating of the pedestal is negligible, the telescope shutter remained closed during data acquisition.
This setup effectively eliminates contributions from external phenomena such as signals of an EAS, lightning, muons, or airplanes.
Approximately $4.174$ million traces were collected per PMT, resulting in a total of approximately $16.698$ million traces.

\begin{figure}[htb!]
    \centering
    \includegraphics[width=0.45\linewidth]{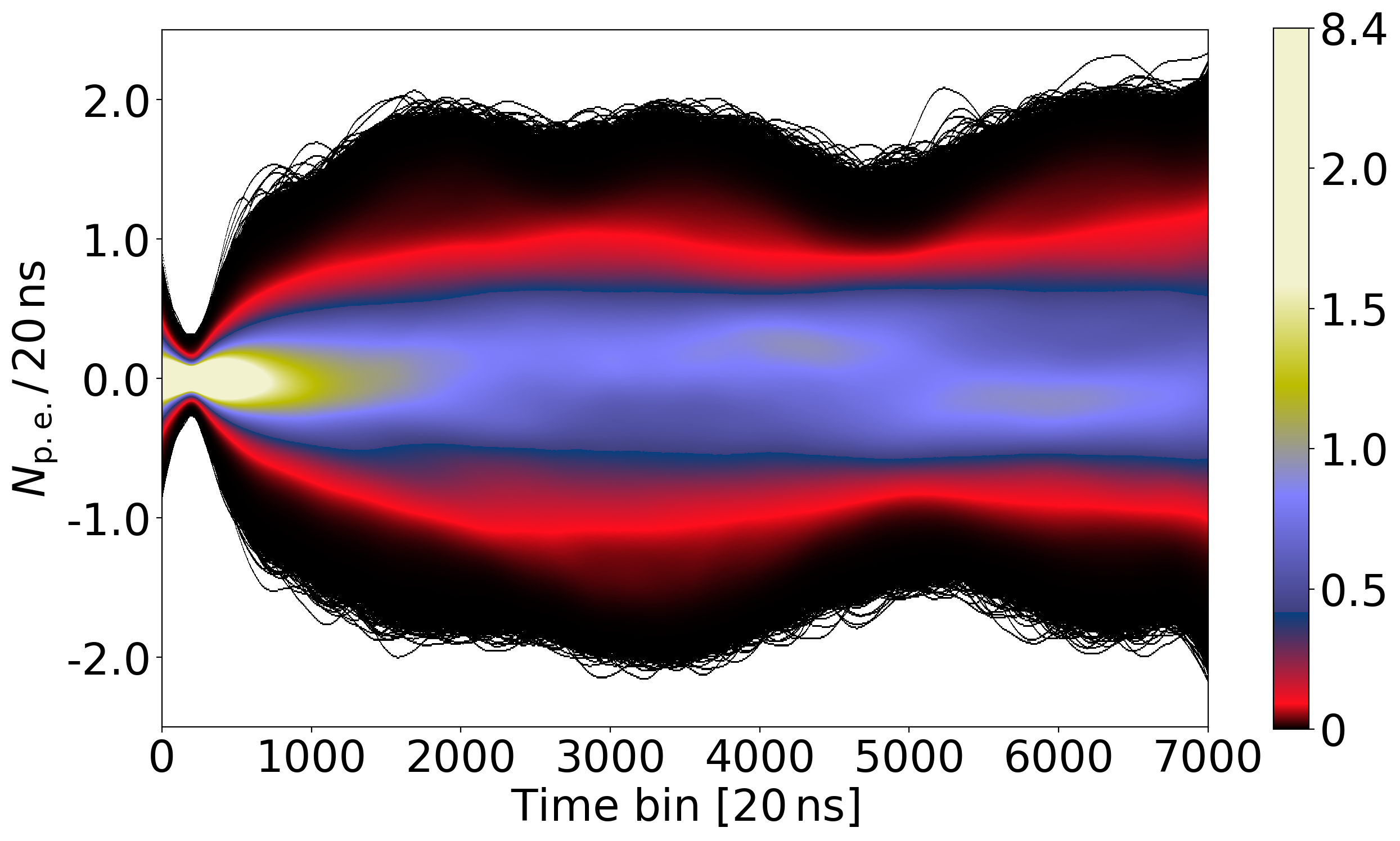}
    \caption{Two-dimensional histogram of all baselines extracted from pedestal data acquired with the telescope shutter closed.
    The values, which are colored according to the colour bar on the right, are normalized to represent a PDF for each bin.
    Note that values above two have the same colour to show details for lower values.}
    \label{fig:baselines}
\end{figure}

The baseline of each trace was extracted using a two-step filtering process: an MA filter with a window length of $513$ bins, followed by a third-order Savitzky-Golay filter \cite{Savitzky1964} with the same window length.
Examples of individual traces with their extracted baselines have already been shown in Fig. \ref{fig:motivation_01}.
All extracted baselines (in total $16,\!697,\!656$) are visualized in Fig.~\ref{fig:baselines} as a two-dimensional histogram that is normalized to represent a probability density function (PDF) for each time bin. 
The floating nature of the baselines is clearly visible.
The ``narrow bottleneck'' observed in the first few hundred bins is a result of shifting the pedestal to zero during data acquisition and does not affect the threshold calculations in the following section.

\subsection{Threshold calculation}
\label{subsec:thresholds}
The standard procedure during data acquisition is to collect as much data as possible, typically constrained by storage capacity and the limitations of the readout electronics.
Trigger thresholds should be low enough to enable the detection of weak signals, while still avoiding excessive triggering from the NB.
To achieve this balance, thresholds are calculated to maintain a trigger rate of $1.25\,\unit{Hz}$ for the NB per individual filter window and PMT.
The NB is modelled as a sequence of $7000$ bins drawn from $\mathcal{N}(\mu_{\mathrm{NB}},\,\sigma_{\mathrm{NB}}^{2})$ with $\mu_{\mathrm{NB}} = 0$ and $\sigma_{\mathrm{NB}} \in [1.0, 5.0] \, N_{\mathrm{p.e.}}/20\,\unit{ns}$, reflecting varying night-sky conditions.
A baseline is added to each NB trace, resulting in approximately $16.698$ million traces for various $\sigma_{\mathrm{NB}}$ that realistically represent the behavior of the floating pedestal of a single PMT.
SNR values are then computed for all traces across different triggering algorithms, filter window lengths $w_{\mathrm{filt}}$, and standard deviations $\sigma_{\mathrm{NB}}$. 

For a given trace (NB + baseline), SNR values cannot be calculated for every bin, as data before and after the test position~$P_\mathrm{test}$ are required (see Fig. \ref{fig:alg_scheme}).
Therefore, the SNR is computed over the interval from $w_{\mathrm{std}} + w_{\mathrm{ma}} + w_{\mathrm{nan}}$ to $7000 - \frac{w_{\mathrm{filt}}-1}{2}$.
The resulting number of SNR values per trace, denoted as $n_{\mathrm{SNR}}$, ranges from $3983$ to $4171$ bins, depending on the filter window length.
Specifically, $n_{\mathrm{SNR}} = 4171$ for $w_{\mathrm{filt}} = 25$ and $n_{\mathrm{SNR}} = 3983$ for $w_{\mathrm{filt}} = 401$.
Since one bin represents $20\,\unit{ns}$, this corresponds to a time period of $20 \cdot n_{\mathrm{SNR}}\,\unit{ns}$ for a single trace.
As a result, the analysis over all floating baselines covers more than $1300 \, \unit{s}$ for all filter window lengths (calculated as the total number of baselines multiplied by $\frac{20}{10^9} \cdot n_{\mathrm{SNR}}$). 
For instance, for $w_{\mathrm{filt}} = 25$, the total duration is approximately $1393 \,\unit{s}$, while for $w_{\mathrm{filt}} = 401$, it is about $1330 \,\unit{s}$.

To determine thresholds that maintain a specified trigger rate for a given $w_{\mathrm{filt}}$, the maximum SNR value is extracted from each trace.
These maxima are then sorted and the $n-$th largest value is selected, where $n$ is determined by the total duration of analysed baselines and the desired trigger rate.
For example, for $w_{\mathrm{filt}} = 401$ and a trigger rate of $1.25\,\unit{Hz}$, $n$ is set to $1330 \cdot 1.25 \approx 1662$.
The $1662$-nd largest maximum SNR among all traces is then used as the threshold.
Applying this threshold yields exactly $1662$ detected events over $1330\,\unit{s}$, resulting in the intended trigger rate of $1.25\,\unit{Hz}$.
This procedure is repeated for all combinations of filter window lengths, triggering algorithms, and standard deviations $\sigma_{\mathrm{NB}}$.

\begin{figure}[ht!]
    \centering
    \includegraphics[width=0.75\linewidth]{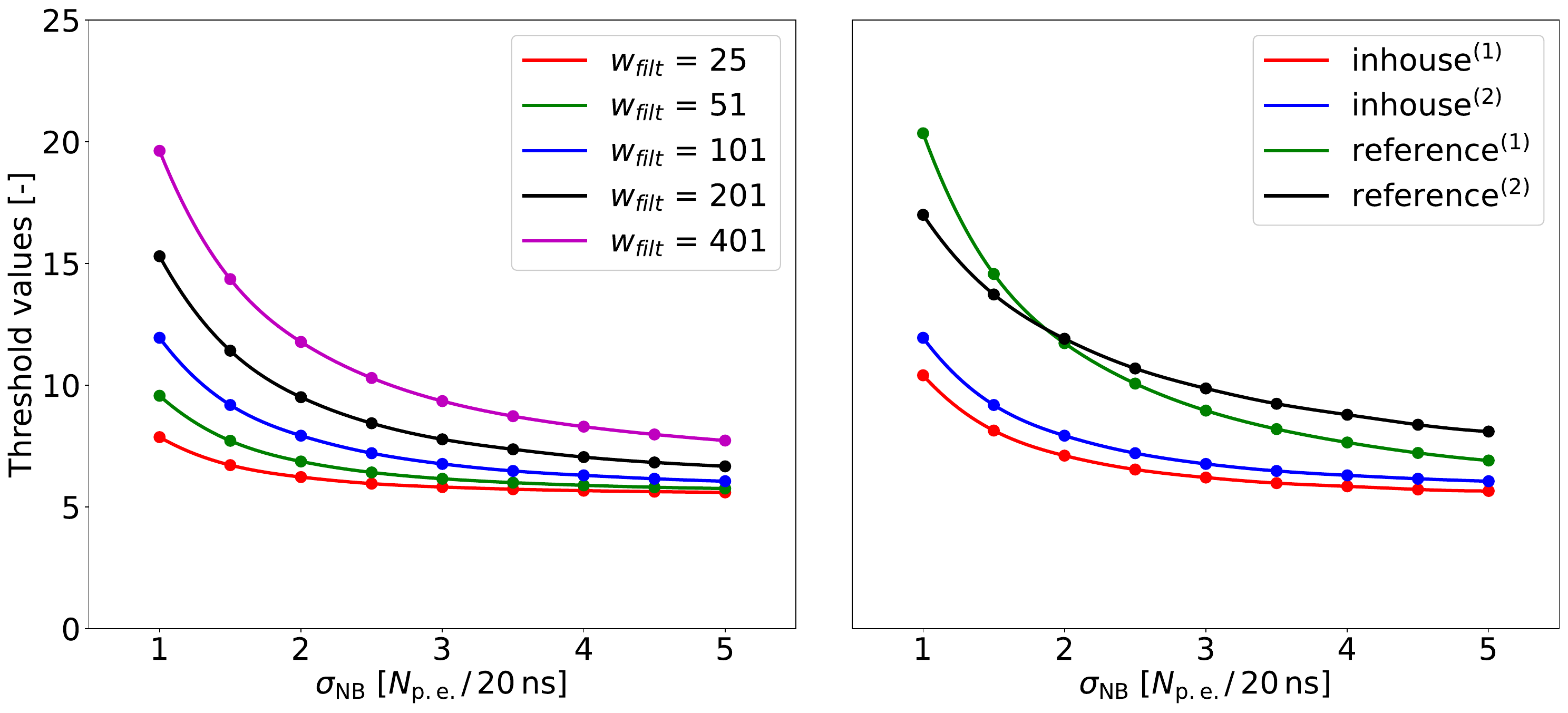}
    \caption{Threshold values required to maintain a trigger rate of $1.25\,\unit{Hz}$ for a single PMT and $w_{\mathrm{filt}}$.
    Left panel: Threshold values for the inhouse$^{(2)}$ algorithm across various $\sigma_{\mathrm{NB}}$ and $w_{\mathrm{filt}}$.
    Right panel: Comparison of threshold values for different algorithms using $w_{\mathrm{filt}} = 101$ for various $\sigma_{\mathrm{NB}}$.}
    \label{fig:thresholds}
\end{figure}

Threshold values are computed for $\sigma_{\mathrm{NB}} \in [1.0, 5.0] \, N_{\mathrm{p.e.}}/20\,\unit{ns}$ in steps of $0.5 \, N_{\mathrm{p.e.}}/20\,\unit{ns}$, and intermediate values are interpolated using a cubic spline.
The obtained thresholds for the inhouse$^{(2)}$ algorithm across various standard deviations~$\sigma_{\mathrm{NB}}$ and filter window lengths are shown in the left panel of Fig. \ref{fig:thresholds},
while the right panel compares the threshold values of the in-house and reference algorithms for different standard deviations $\sigma_{\mathrm{NB}}$ at a fixed filter window length of $w_{\mathrm{filt}} = 101$.
A strong dependence of the threshold values on both $\sigma_{\mathrm{NB}}$ and $w_{\mathrm{filt}}$ is observed for all algorithms, which is a direct consequence of the floating baselines.  
This is because the baseline is unaffected by changes in $\sigma_{\mathrm{NB}}$; therefore, its influence becomes more pronounced at lower~$\sigma_{\mathrm{NB}}$.
Moreover, increasing the filter window length $w_{\mathrm{filt}}$ does not remove the floating baseline, but significantly reduces the standard deviation of the filtered trace, which in turn raises the threshold values. 
When the NB traces are generated without a floating baseline, the threshold values remain nearly constant (see left panels of Fig. \ref{fig:app:thresholds} in Appendix~\ref{appendix:thrsh_calc}).

It is important to note that the comparison in the right panel of Fig. \ref{fig:thresholds} does not indicate which algorithm is the most suitable for data acquisition. 
A lower threshold does not necessarily imply a better signal detection capability, as each algorithm defines the SNR metric differently.
For completeness, threshold values for all algorithms are illustrated in the right panels of Fig. \ref{fig:app:thresholds} in Appendix \ref{appendix:thrsh_calc}.

\section{Results}
Section \ref{subsec:sims} performs the comparison of trigger methods using Monte Carlo simulations of EASs, while the analysis of UHECR events detected by the FAST telescope is presented in Sect. \ref{subsec:uhecr}.
Note that the aim of the following analyses is not to compare FAST with the Auger and TA experiments. 
Instead, we apply different triggering algorithms given by Eq. (\ref{eq:snr_definitions}), some based on those used in Auger and TA (reference algorithms), and others developed specifically for FAST (in-house algorithms), in order to identify the most effective triggering method for FAST.

\subsection{UHECR simulations}
\label{subsec:sims}
\begin{figure}[ht!]
    \centering
    \includegraphics[width=0.35\linewidth]{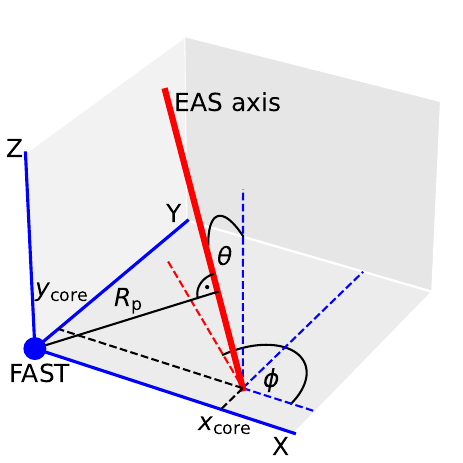}
    \caption{The geometric parameters of the EAS. 
    The impact parameter $R_\mathrm{p}$ is defined as the closest distance between the reconstructed shower axis and the detector, and it is used in the UHECR analysis.}
    \label{fig:geometry}
\end{figure}

The simulations of EASs for a single FAST telescope are generated using the FAST framework \cite{Malacari2020}.
In these simulations, the FAST telescope is placed at coordinates [$0$, $0$] and its pointing direction is along the $y-$axis.
Each simulation is characterized by six parameters: primary particle energy~$E$, the depth of the EAS maximum $X_{\mathrm{max}}$, zenith angle $\theta$, azimuth angle $\phi$, and core position [$x_{\mathrm{core}}$, $y_{\mathrm{core}}$], where the latter denotes the impact location on the ground.
The geometric parameters of the EAS are illustrated in Fig. \ref{fig:geometry}, with the exception of $X_{\mathrm{max}}$, which is discussed, e.g., in \cite{Abbasi2023}.
All parameters are sampled from continuous uniform distributions over predefined intervals,
except $X_{\mathrm{max}}$, for which a more realistic approach is adopted to ensure physically relevant parameter combinations. 
$X_{\mathrm{max}}$ values are drawn from energy-dependent PDFs derived from CONEX simulations \cite{Bergmann2007} for protons as primary particles.
These density functions are illustrated in Fig. \ref{fig:app:pdfs} in Appendix \ref{appendix:pdfs}. 
The parameter distributions are as follows:
\begin{itemize}
    \item $E$ [eV]: drawn from a continuous uniform distribution in $\log_{10}$ scale over the interval [$17.5, 20.5$].
    \item $X_{\mathrm{max}}$ [g/cm$^2$]: drawn from energy-dependent PDFs for UHECRs with protons as primary particles.
    \item $\phi$ [degree]: drawn from a continuous uniform distribution over [$-180, 180$].
    \item $\theta$ [degree]: drawn from a continuous uniform distribution over [$0, 90$].
    \item $x_{\mathrm{core}}$ [m]: drawn from a continuous uniform distribution over [$-12500, 12500$].
    \item $y_{\mathrm{core}}$ [m]: drawn from a continuous uniform distribution over [$0, 25000$].
\end{itemize}

Moreover, only core positions within a radius of $25000\,\unit{m}$ from the telescope are considered, and only if they lie within the field of view of the FAST telescope ($30\degree$ in azimuth) extended by $15\degree$ on each side, i.e., a total of $60\degree$ in azimuth directly in front of the telescope.
This accounts for the fact that the shower core may lie outside the field of view of the FAST telescope, while the corresponding EAS is still observable in the atmosphere above.
Note that a uniform energy distribution is adopted to ensure a sufficient range of signals with varying maximum amplitudes for additional analysis. 
Therefore, matching the actual UHECR flux-energy dependence is unnecessary.
A total of $1.5$ million simulations were generated.
Events where the maximum of the signal across all four PMTs was lower than $0.1 \, N_{\mathrm{p.e.}}/20\,\unit{ns}$ were excluded, 
as they contain negligible or no detectable signal.
This results in a total of $834,\!973$ usable simulations for the comparison.

\begin{figure}[ht!]
    \centering
    \includegraphics[width=0.70\linewidth]{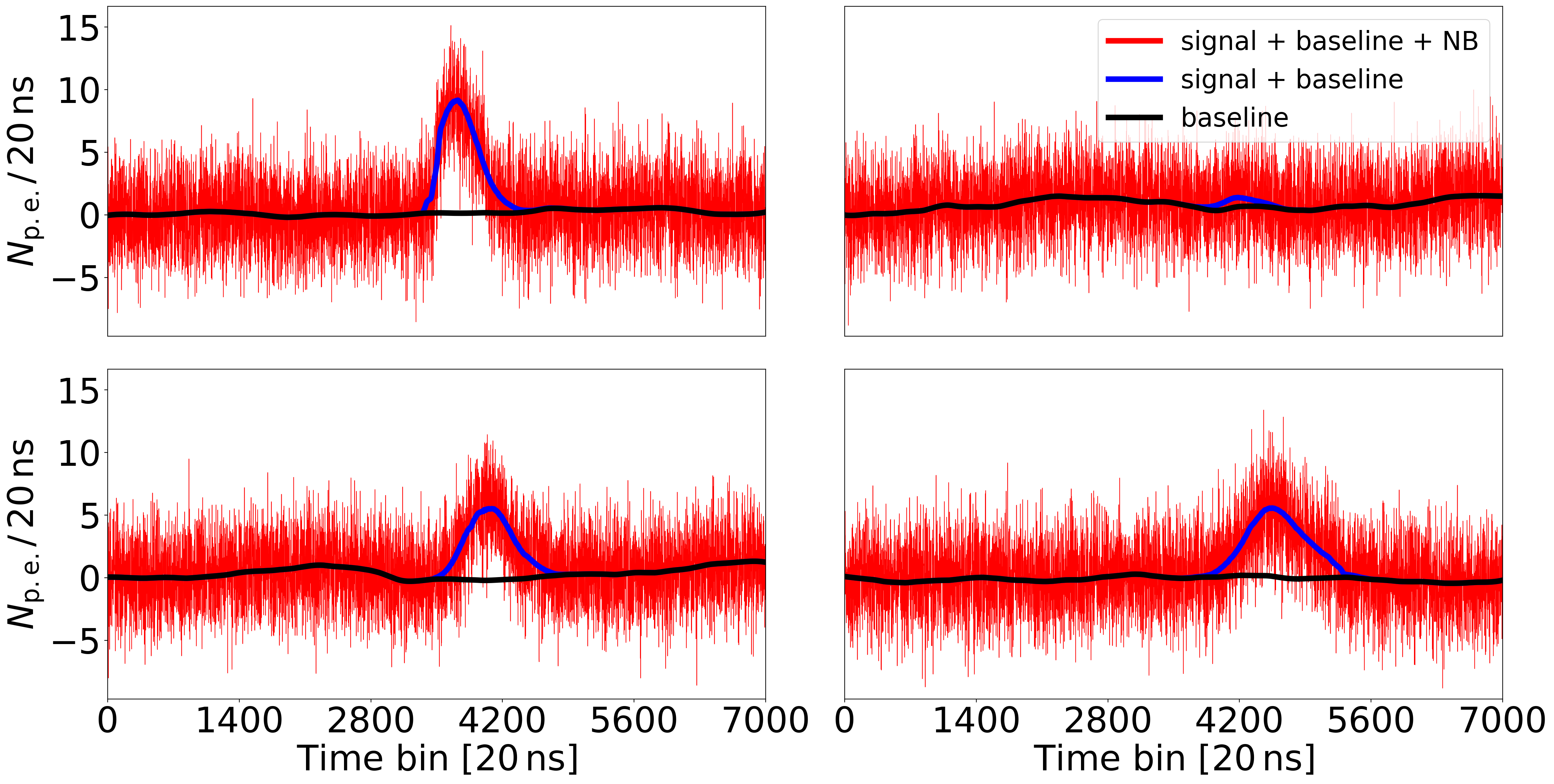}
    \caption{Example of a single simulated event for the FAST telescope consisting of four PMTs, generated using the following parameters:
    $E = 100\,\unit{EeV}$, $X_{\mathrm{max}} = 813 \, \unit{g/cm}^2$, $\phi= -177 \degree$, $\theta=52 \degree$, $x_{\mathrm{core}} = 4168 \,\unit{m}$, and $y_{\mathrm{core}} = 18010 \,\unit{m}$. 
    The PMT layout reflects the sky view: the upper two PMTs observe elevations above $15\degree$, while the lower two observe below $15\degree$. 
    The NB is drawn from $\mathcal{N}(\mu_{\mathrm{NB}},\,\sigma_{\mathrm{NB}}^{2})$ with $\mu_{\mathrm{NB}} = 0$ and $\sigma_{\mathrm{NB}}=~2.5\,N_{\mathrm{p.e.}}/20\,\unit{ns}$. 
    The final event, shown in red, includes the simulated signal, floating baseline, and the NB.}
    \label{fig:sim}
\end{figure}

For each simulated trace, an NB is added from $\mathcal{N}(\mu_{\mathrm{NB}},\,\sigma_{\mathrm{NB}}^{2})$, along with a randomly selected floating baseline from the dataset visualized in Fig.~\ref{fig:baselines}.
An example of a single simulated event as observed by the FAST telescope is presented in Fig. \ref{fig:sim}, where each subplot corresponds to one PMT.
In this example, for each PMT, an NB with $\sigma_{\mathrm{NB}} = 2.5\,N_{\mathrm{p.e.}}/20\,\unit{ns}$ and a random baseline (shown as a solid black line) are added to the signal from the simulated event. 
The resulting event, which includes signal from the simulated event, the floating baseline and the NB, is shown in red.

For the given triggering algorithm, the simulation is considered to contain a signal from an EAS if at least one PMT passes.
To evaluate whether a PMT passes the trigger algorithm, the maximum SNR values are calculated for the five different filter window lengths $w_{\mathrm{filt}}$ described in Sect. \ref{subsec:algorithms}.
Each of these five values per PMT is then compared to its corresponding thresholds derived from the previous NB analysis. 
A PMT is said to detect a signal if at least one of these five SNR values exceeds its respective threshold.

\begin{figure}[ht!]
    \centering
    \includegraphics[width=0.37\linewidth]{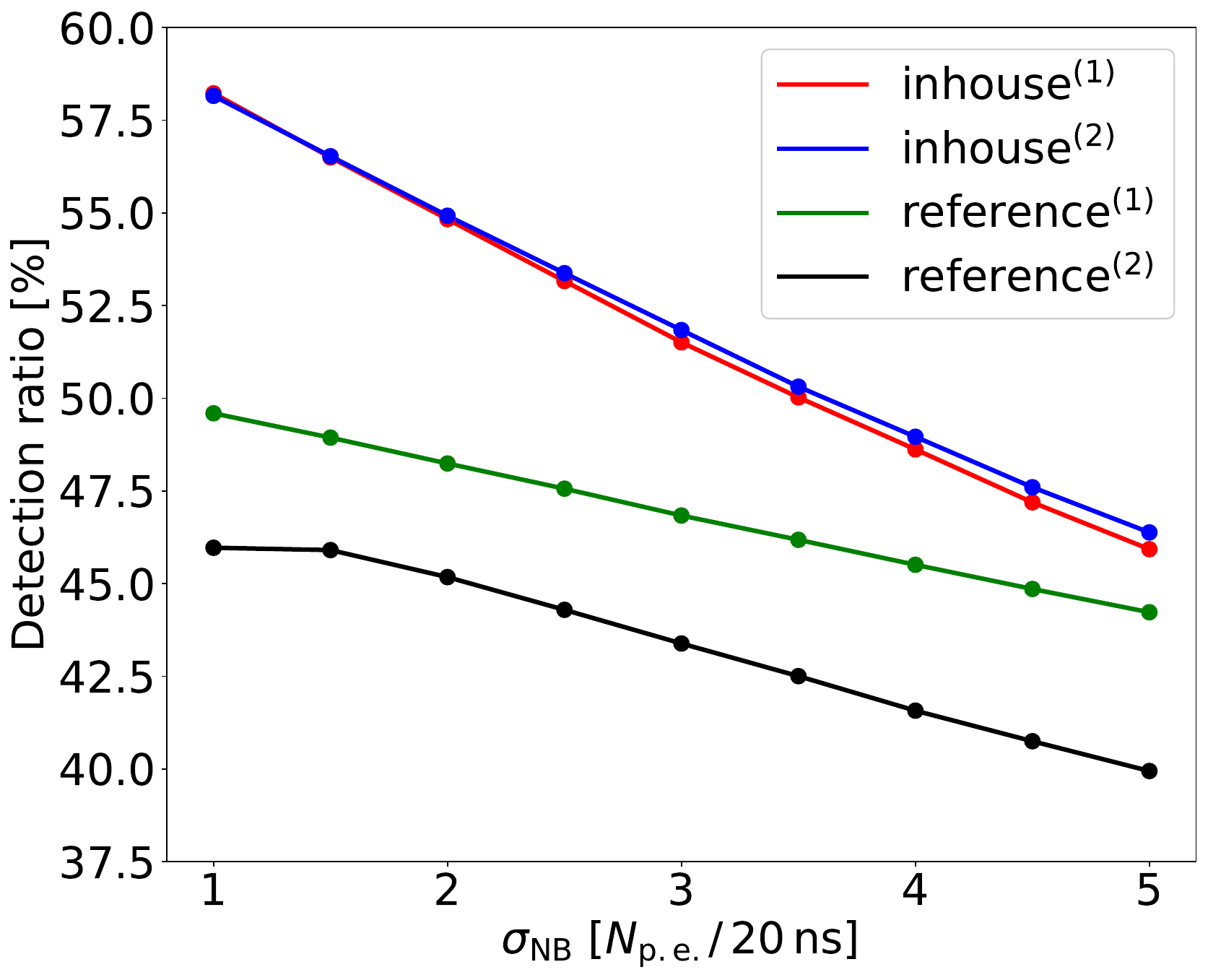}
    \caption{Detection ratio of UHECR simulations for a single FAST telescope and for various triggering algorithms as a function of $\sigma_{\mathrm{NB}}$.
    The in-house algorithms clearly outperform the reference algorithms across all noise background levels.
    See the text for further details.}
    \label{fig:sim_eval_01}
\end{figure}

Figure \,\ref{fig:sim_eval_01} shows the detection ratio of UHECR simulations for a single FAST telescope, comparing different triggering algorithms across a range of standard deviations $\sigma_{\mathrm{NB}} \in [1.0, 5.0] \, N_{\mathrm{p.e.}}/20\,\unit{ns}$. 
This range of $\sigma_{\mathrm{NB}}$ is sufficient to cover all night-sky conditions encountered during measurements.
The detection ratio is defined as the fraction of simulations containing a signal relative to the total number of usable simulations.
While the in-house algorithms exhibit a similar performance across different $\sigma_{\mathrm{NB}}$ values, the reference algorithms show lower detection ratios. 
Moreover, a comparison between the reference algorithms reveals that using $\sigma(\mathrm{trace}_{\mathrm{filt}})$ substantially reduces the performance of the reference$^{(2)}$ algorithm.
This highlights an important insight: when estimating $\sigma(\mathrm{trace}_{\mathrm{filt}})$, it is preferable to use the standard deviation of the unfiltered trace $\sigma(\mathrm{trace})$, as described by Eq.~(\ref{eq:sigma}).

It is evident that the performance of all algorithms decreases as $\sigma_{\mathrm{NB}}$ increases, due to the lesser significance of the signal compared to the NB.
However, this trend is less pronounced for the reference$^{(1)}$ algorithm, whose performance becomes increasingly comparable to that of the in-house algorithms at higher $\sigma_{\mathrm{NB}}$ values. 
This is expected as the influence of floating baselines becomes smaller with increasing NB, making the baseline estimation less critical.
As a result, when the effect of floating baselines is sufficiently small (e.g., by increasing $\sigma_{\mathrm{NB}}$), the reference$^{(1)}$ algorithm can outperform the in-house methods.
In the limiting case where floating baselines are negligible, the reference$^{(1)}$ algorithm always provides better performance.
This behavior directly results from the estimation of $\mathrm{trace}_{\mathrm{mean}}(P_{\mathrm{ma}})$ in Eqs. (\ref{eq:snr1}) and (\ref{eq:snr2}), which should ideally be zero in the absence of a floating baseline.
This limiting scenario, where no floating baseline is present (i.e., corresponding to very high~$\sigma_{\mathrm{NB}}$), is illustrated in Fig. \ref{fig:app:detection_ratio} in Appendix \ref{appendix:detection_ratio}, which shows the detection ratio using thresholds derived without including floating baselines (see left panels of Fig. \ref{fig:app:thresholds}).
The necessity of the in-house algorithms in the FAST experiment thus arises from the baseline fluctuations as their size is comparable to $\sigma_{\mathrm{NB}}$.
These fluctuations impact the SNR calculation, making the baseline estimation essential for reliable triggering.

\begin{figure}[ht!]
    \centering
    \includegraphics[width=0.74\linewidth]{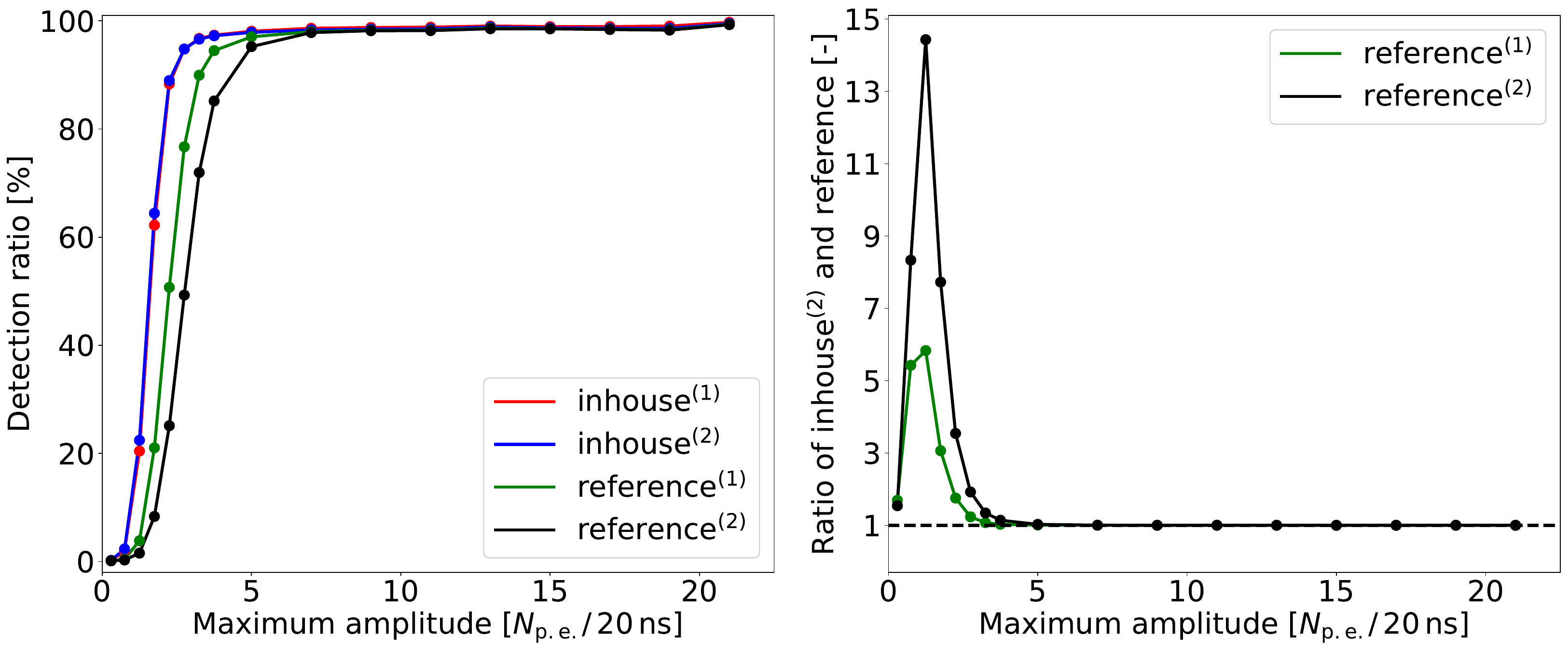}
    \caption{Detection ratio of simulations across all triggering algorithms as a function of maximum amplitude (left panel).
    The in-house algorithms clearly outperform the reference algorithms at lower maximum amplitudes.
    The details are well illustrated in the right panel, where the ratio of detected events between the inhouse$^{(2)}$ and reference algorithms is shown.
    For both panels, all events with a maximum amplitude exceeding $20 \, N_{\mathrm{p.e.}}/20\,\unit{ns}$ are grouped into the final interval ($> 20\, N_{\mathrm{p.e.}}/20\,\unit{ns}$), as finer resolution is not informative in this range.}
    \label{fig:sim_eval_02}
\end{figure}

Note that the simulation-based analysis is intended purely for comparing different triggering algorithms. 
Accordingly, the detection ratios in Fig. \ref{fig:sim_eval_01} do not represent the realistic sensitivity of FAST.
Such a study would require accounting for the realistic UHECR flux-energy dependence.
However, the use of a uniform energy distribution is beneficial here, as it allows a more detailed comparison of triggering algorithms, particularly highlighting the specific conditions under which the in-house algorithms outperform the reference algorithms. 
Since the maximum signal amplitude of an event scales with its energy, the maximum amplitude across all PMTs provides a meaningful metric for this analysis.
The chosen energy distribution ensures a sufficient number of simulated EASs with varying maximum amplitudes.
The left panel of Fig. \ref{fig:sim_eval_02} shows the detection ratio as a function of the maximum amplitude for each algorithm, using $\sigma_{\mathrm{NB}} = 2.5 \, N_{\mathrm{p.e.}}/20\,\unit{ns}$. 
Although this specific value is used for the comparison, the same trend is observed across all tested values of $\sigma_{\mathrm{NB}} \in [1.0, 5.0] \, N_{\mathrm{p.e.}}/20\,\unit{ns}$.
Note that $\sigma_{\mathrm{NB}} = 2.5 \, N_{\mathrm{p.e.}}/20\,\unit{ns}$ reflects a typical NB during data acquisition.
During measurements taken between 4.7.2022 and 23.10.2022, the average $\sigma_{\mathrm{NB}}$ was approximately $2.29 \, N_{\mathrm{p.e.}}/20\,\unit{ns}$.

As expected, events with sufficiently high maximum amplitude are reliably detected by all algorithms.
However, the performance gap becomes significant at lower amplitudes.
This is clearly illustrated in the right panel of Fig.~\ref{fig:sim_eval_02}, which shows the ratio of detected events between the inhouse$^{(2)}$ algorithm and both reference algorithms.
For example, in the amplitude range of $1.5-2.0 \, N_{\mathrm{p.e.}}/20\,\unit{ns}$, inhouse$^{(2)}$ achieves a detection ratio over three times higher than reference$^{(1)}$,
and in the range $1.0-1.5 \, N_{\mathrm{p.e.}}/20\,\unit{ns}$, the ratio increases to nearly six.
As the maximum amplitude exceeds $4 \, N_{\mathrm{p.e.}}/20\,\unit{ns}$, the ratio converges toward unity, indicating similar performance across all algorithms. 
This is not surprising: if the signal is strong enough, it is detected regardless of the algorithm. 
It also explains why there is no greater difference between the in-house and reference algorithms in Fig.~\ref{fig:sim_eval_01}, since more than $38\%$ of the simulations used for the analysis have the maximum amplitudes above $4\,N_{\mathrm{p.e.}}/20\,\unit{ns}$. 
For weaker signals, however, the in-house algorithms clearly dominate, which is a crucial feature for FAST, as most UHECRs detected by FAST have relatively low maximum amplitudes.
This is further demonstrated for UHECR events detected by the FAST telescope in Sect. \ref{subsec:uhecr}.

Figure \ref{fig:sim_eval_02} shows that the in-house algorithms exhibit comparable performance, which is consistent with the results in Fig.~\ref{fig:sim_eval_01}.
While inhouse$^{(2)}$ performs slightly better at lower maximum amplitudes, the overall performance difference is negligible.
This makes it difficult to determine which algorithm is more suitable for data acquisition.
A more detailed evaluation is thus provided in Sect. \ref{subsec:disc:ma} in the Discussion section.

\subsection{UHECR events detected by the FAST telescope}
\label{subsec:uhecr}
The algorithms developed in this paper are intended for triggering at the hardware level.
To demonstrate their suitability, they are applied to data from the FAST telescope obtained via the external trigger from Auger.
A total of $1463$ candidate events in time coincidence with Auger were collected by the single FAST telescope between 4.7.2022 and 24.10.2022.
Each event consists of four traces, each containing $5000\,$bins with each bin corresponding to $20\,\unit{ns}$.
Some events have energies below the defined UHECR threshold of $0.1\,\unit{EeV}$ \cite{Coleman2023}; however, they are included in the analysis for completeness.
Note that UHECRs that pass the trigger algorithm would be acquired during nighttime observations.
In the following, such events are referred to as being detected by the respective trigger algorithm.

The events were stored such that $1720$ bins before the timing of the external trigger were saved, so a potential signal begins at bin $1720$.
To ensure a sufficient amount of data before the signal for applying the developed algorithms, a minor adjustment was made to the dataset:
$1280\,$bins from the end of each trace were shifted to the beginning, relocating the signal start position to bin $3000$.
These bins are not expected to contain any signal. 
This results in calculating the standard deviation in Eq. (\ref{eq:snr_definitions}) from the region which now includes bins from both the original beginning and end of the trace.
However, this modification has no effect on the triggering algorithms, since the characteristics of the NB remain consistent before and after the signal.

The triggering algorithms are applied over the bin interval $3000-3350$, where the signal is expected to occur.
In simulations, $\sigma_{\mathrm{NB}}$ is predefined and thresholds are set accordingly, whereas for candidate events, it must be estimated.
For each trace, $\sigma_{\mathrm{NB}}$ is approximated using the first $2048$ bins.
An event is then considered to contain a signal if at least two PMTs register a signal above the respective threshold.
This requirement is similar to the second-level trigger used in the Auger and TA experiments.
Although this condition is more strict than that applied in the simulations, it effectively suppresses triggers from noise fluctuations or electronics artifacts, particularly when using low thresholds to maintain a maximum $6.25\,\unit{Hz}$ trigger rate per PMT.
However, given that the field of view of a single PMT is $15\degree \times 15\degree$ in azimuth and elevation, it is typical for even strong signals to appear in only one PMT.
To ensure that such events are not discarded, an additional condition is applied:
an event is also considered to contain a signal from an EAS if at least one PMT registers a strong signal.
A strong signal is defined as one whose SNR exceeds $1.5$ times the trigger threshold. 
In the case of stereo observation, using multiple FAST telescopes arranged in a triangular array, a stereo trigger can be implemented by requiring coincidences between individual detectors.
The implementation of a stereo trigger scheme is discussed in more detail in the Discussion section (Sect. \ref{subsec:disc:trigger}). 

Table \ref{tab:uhecr} summarizes the number of UHECR events detected by each triggering algorithm.
An event is classified as a ``$\geq 1\,$PMT w. strong signal'' category if at least one PMT exceeds the strong signal condition, regardless of the number of detected signals in the remaining PMTs. 
The second category, ``$\geq 2\,$PMTs w. signal'', includes events where two or more PMTs exceed the trigger threshold, but none exceed $1.5$ times the threshold.
This results in two disjoint sets of detected events.
Unsurprisingly, the two in-house algorithms perform almost identically.
However, the performance gap between the in-house algorithms and the reference algorithms is much larger than in simulations (see Fig.~\ref{fig:sim_eval_01}).
This is because the real candidate events are generally weaker than those in the simulation dataset, which was based on a uniform energy distribution, resulting in a large number of very high energy events with high-amplitude signals.
Since the in-house algorithms significantly outperform the reference algorithms in the case of weaker signals (see Fig. \ref{fig:sim_eval_02}), the observed difference is increased.

\begin{table}[h!]
    \caption{Number of UHECR events detected by different triggering algorithms.
    The category ``$\geq 1\,$PMT w. strong signal'' denotes events with at least one PMT exceeding $1.5$ times the trigger threshold, while ``$\geq 2\,$PMTs w. signal'' requires two or more PMTs above the threshold, but none above $1.5$ times the trigger threshold.
    A total of $1463$ candidate events were used for the analysis.}
    \centering
    \begin{tabular}{l||c|c|c|}
    algorithm        & $\geq 1\,$PMT w. strong signal & $\geq 2\,$PMTs w. signal & total \\
    \hline
    \hline
    inhouse$^{(1)}$  & $209$  & $60$  & $269$  \\
    inhouse$^{(2)}$  & $202$  & $66$  & $268$  \\
    reference$^{(1)}$ & $139$  & $24$  & $163$  \\
    reference$^{(2)}$ & $74$   & $3$   & $77$   \\
    \end{tabular}
    \label{tab:uhecr}
\end{table}

\begin{figure}[htb!]
    \centering
    \includegraphics[width=0.73\linewidth]{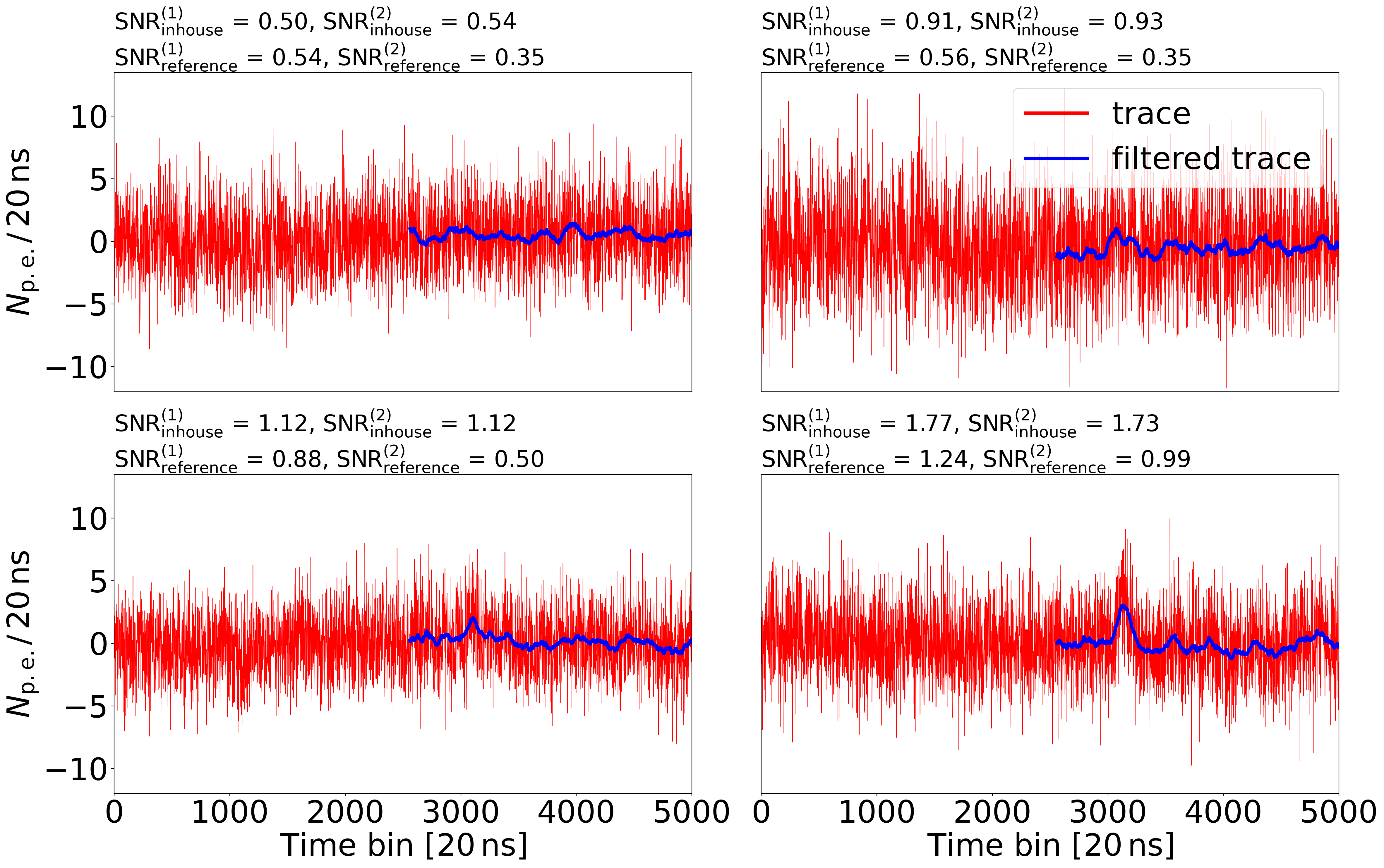}
    \caption{Example of a UHECR event classified as a strong signal by the in-house algorithms but not by the reference algorithms. 
    The PMT layout reflects the sky view: the upper two PMTs observe elevations above $15\degree$, while the lower two observe below $15\degree$.
    The maximum normalized SNR values in each panel title illustrate the increased sensitivity to signals when using the in-house algorithms, particularly in the bottom PMTs.}
    \label{fig:uhecr}
\end{figure}

An example of a UHECR event detected by both in-house algorithms but missed by the reference algorithms is shown in Fig. \ref{fig:uhecr}.
The PMT layout again corresponds to the sky view as in Fig. \ref{fig:sim}.
The filtered traces, shown in blue, were obtained by applying an MA filter with a window size of $w_{\mathrm{filt}} = 101\,$bins.
The reconstructed energy $E$ and the impact parameter~$R_\mathrm{p}$ (the closest distance between the reconstructed shower axis and detector defined in Fig. \ref{fig:geometry}) are $E = 3.82\,\unit{EeV}$ and $R_\mathrm{p} = 3.13\,\unit{km}$, as provided by a preliminary hybrid reconstruction from Auger.
For each filter window length $w_{\mathrm{filt}}$, the normalized SNR is calculated as the ratio of the computed SNR to its respective threshold. 
For each algorithm, the maximum normalized SNR across all values of $w_{\mathrm{filt}}$ is shown in the title of each PMT panel. 
A strong signal is observed in the bottom-right PMT for both in-house algorithms, as they yield values exceeding $1.70$ times the trigger threshold. 
In contrast, the reference$^{(1)}$ algorithm reaches a maximum normalized SNR of only $1.24$, and the reference$^{(2)}$ algorithm does not reach the threshold at all for any $w_{\mathrm{filt}}$, as the maximum normalized SNR is below unity.
Moreover, a weak signal is observed in the bottom-left PMT but again only for the in-house algorithms.

\begin{figure}[ht!]
    \centering
    \includegraphics[width=0.63\linewidth]{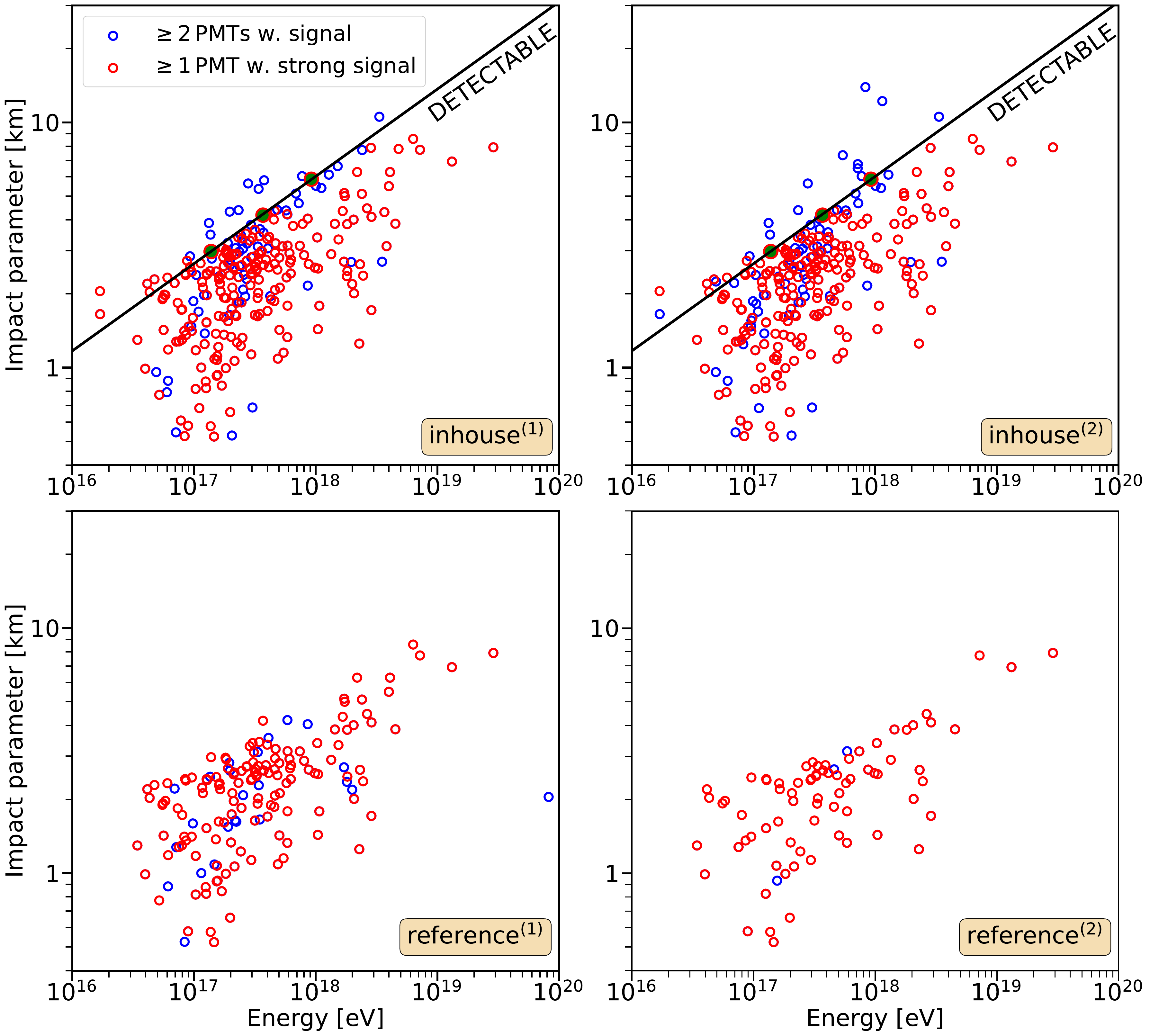}
    \caption{Impact parameter as a function of energy for events detected by the in-house and reference triggering algorithms. 
    The two classes of detected events (see Table \ref{tab:uhecr}) are distinguished by colour.
    Reconstructed parameters are obtained from a preliminary hybrid reconstruction from Auger available in \cite{Bradfield2025}.
    The same black line in both upper panels is fitted to the extreme data points of strong signal (red points) with an energy above~$0.1\,\unit{EeV}$. This line roughly indicates the maximum detectable shower distance for a given energy in the case of the in-house methods.
    The extreme data points used for the fit are marked in green.}
    \label{fig:uhecr:erp}
\end{figure}

\begin{figure}[ht!]
    \centering
    \includegraphics[width=0.63\linewidth]{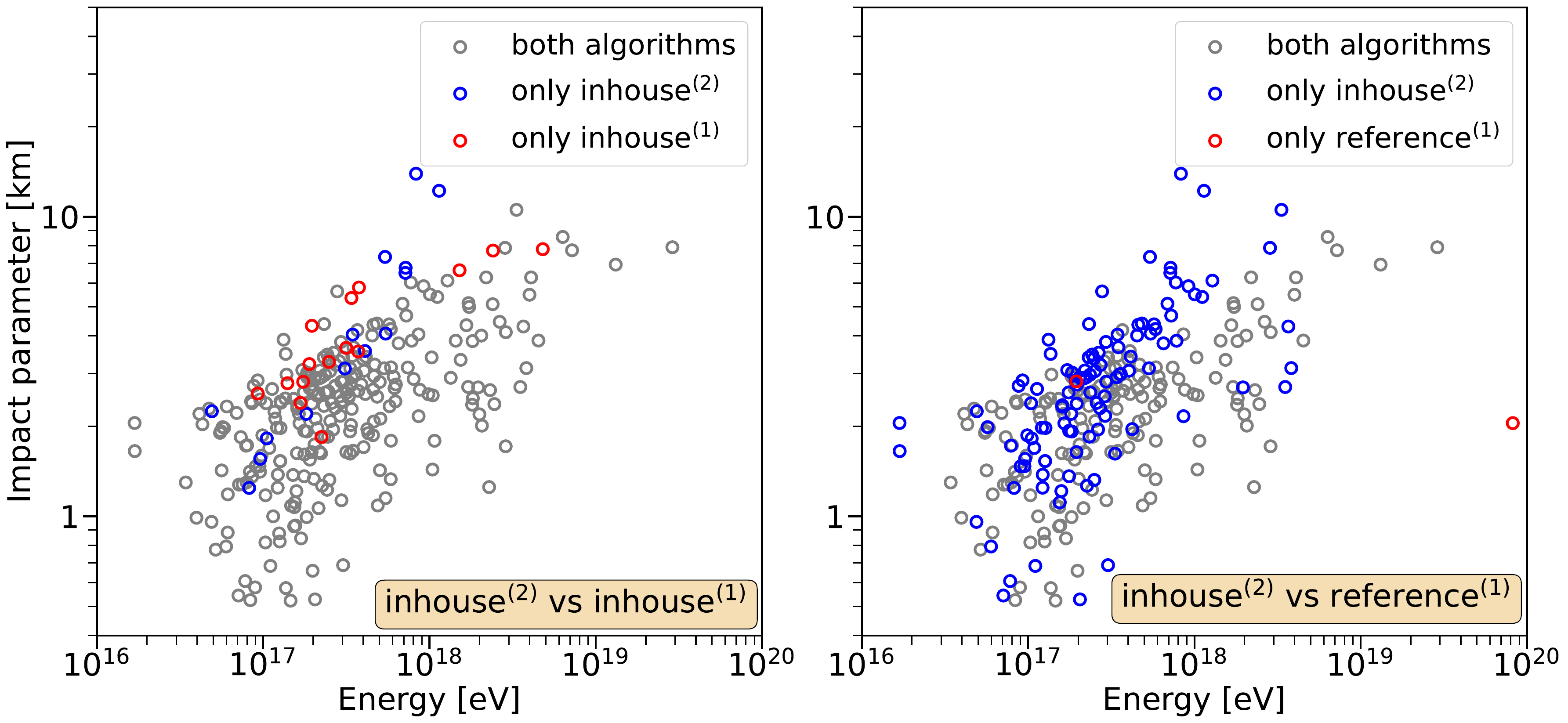}
    \caption{Comparison of detected events between inhouse$^{(2)}$ and inhouse$^{(1)}$ algorithms (left panel), and inhouse$^{(2)}$ and reference/TA$^{(1)}$ (right panel).  
    Both event classes from Table \ref{tab:uhecr} are merged for each algorithm.
    Reconstructed parameters are provided by a preliminary hybrid reconstruction from Auger available in \cite{Bradfield2025}.}
    \label{fig:uhecr:erp_compare}
\end{figure}

The relationship between impact parameter and energy (as determined by a preliminary hybrid reconstruction from Auger available in \cite{Bradfield2025}) for events detected by each triggering algorithm is shown in Fig. \ref{fig:uhecr:erp}.
Events classified as strong signals are marked in red, while those satisfying the ``$\geq 2\,$PMTs w. signal'' condition are shown in blue.
To better highlight the difference between algorithms, Fig.~\ref{fig:uhecr:erp_compare} presents a comparison between the two.
The left panel compares inhouse$^{(2)}$ with inhouse$^{(1)}$, while the right panel compares inhouse$^{(2)}$ with reference$^{(1)}$.
In both cases, the dataset for each algorithm includes all detected UHECR events, regardless of whether they are classified as a strong signal or not.
Events detected by both algorithms are marked in grey,
and those detected only by inhouse$^{(2)}$ and the second comparison algorithm (either inhouse$^{(1)}$ or reference$^{(1)}$) are shown in blue and red, respectively.
The left panel confirms that even in this specific energy-radius relationship, the difference between the two in-house algorithms is negligible.
On the contrary, the right panel reveals a markable discrepancy between inhouse$^{(2)}$ and the reference$^{(1)}$ algorithm. This observation holds for comparison between any in-house and reference algorithm and further demonstrates the limitations of the reference algorithms when applied to FAST data.

The upper panels of Fig. \ref{fig:uhecr:erp} also provide a rough estimate of the FAST telescope sensitivity -- namely, which UHECR events could be detected with the novel internal trigger approach, independent of the Auger experiment.
However, it is important to consider two major biases introduced by the use of the external trigger dataset.
First, the external trigger affects which events are recorded.
Since the Auger and FAST detectors have different characteristics, this may result in a selection bias. 
Second, a preliminary hybrid reconstruction from Auger provides only a preliminary estimation of event parameters, which may lead to inaccuracies.
While these biases do not impact the previous comparisons of trigger algorithms, they clearly affect the sensitivity estimate of the FAST telescope.
Yet, the results still offer a meaningful approximation.

The sensitivity estimate, based only on strong signals (red points) detected by the in-house algorithms with an energy above~$0.1\,\unit{EeV}$, is illustrated by the same black line in both upper panels of Fig. \ref{fig:uhecr:erp}.
This line is fitted to the extreme strong-signal data points marked in green.
It indicates the maximum distance at which an event of a given energy can be detected, but it must be understood as a lower-bound estimate, since only strong signals were used for the linear fit. 
Based on this estimate, events with energies of approximately $60\,\unit{EeV}$ should be detectable at distances up to $\approx 20\,\unit{km}$.
This aligns well with results from the FAST telescope in the northern hemisphere, where estimates indicated that UHECRs detected at the same $20\,\unit{km}$ distance would require a minimum energy of $\approx 40\,\unit{EeV}$~\cite{Malacari2020}.
This small difference can likely be attributed to differences in the event search strategy, the fitting procedure used in Fig. \ref{fig:uhecr:erp}, which considers only strong-signal events, and uncertainties in the parameter reconstructions. 
The consistency between the FAST telescopes in both hemispheres is promising and supports the potential of FAST to help resolve the energy scale difference between the Auger and TA experiments~\cite{Plotko2023}.
The sensitivity analysis is also consistent with a theoretical estimate for the FAST, which predicts that events with energies of $100\,\unit{EeV}$ are detectable at distances up to $25\,\unit{km}$ \cite{Ahlers2025} (see Table~2 in \cite{Ahlers2025}).

\section{Discussion}

\subsection{On the use of MA filters}
\label{subsec:disc:ma}
While the inhouse$^{(1)}$ algorithm applies an FIR filter with a Hamming window, the inhouse$^{(2)}$ uses an MA filter, which is equivalent to an FIR filter with a boxcar window.
Since the MA clearly has a worse frequency response compared to the Hamming window (see Fig. \ref{fig:app:freq} in Appendix \ref{appendix:calib03}), one might expect this to be reflected in the resulting SNR calculation.
Surprisingly, comparisons based on simulations (see Figs. \ref{fig:sim_eval_01} and~\ref{fig:sim_eval_02}) and UHECR events (see Table \ref{tab:uhecr} and Fig. \ref{fig:uhecr:erp_compare}) do not indicate any notable differences.
On the contrary, the inhouse$^{(2)}$ approach appears to be at least as suitable as inhouse$^{(1)}$. 
Therefore, it remains unclear which of the two in-house algorithms is more appropriate for data acquisition.

To investigate this further, we compute the maximum normalized SNR across all values of $w_{\mathrm{filt}}$ for each PMT and simulation event from Sect. \ref{subsec:sims}, assuming $\sigma_\mathrm{NB} = 2.5\, N_{\mathrm{p.e.}}/20\,\unit{ns}$.
Let us remind that the normalized SNR is defined as the ratio of the computed SNR to its respective threshold, and therefore represents how significantly the signal stands out.
For simplicity, the term ``normalized SNR'' will refer to the maximum normalized SNR throughout the following text.
From the full simulation dataset, we select only those events, and corresponding traces of individual PMTs, where the normalized SNR (averaged between the two in-house algorithms) is greater than $0.90$.
This ensures that traces close to the detection threshold are included in the analysis.
Next, we compute the ratio of normalized SNR between inhouse$^{(2)}$ and inhouse$^{(1)}$ for all selected traces.
The average ratio across these traces is approximately $0.968$. 
This indicates that, on average, inhouse$^{(2)}$ achieves a normalized SNR about $3\,\%$ lower than inhouse$^{(1)}$.

\begin{figure}[ht!]
    \centering
    \includegraphics[width=0.39\linewidth]{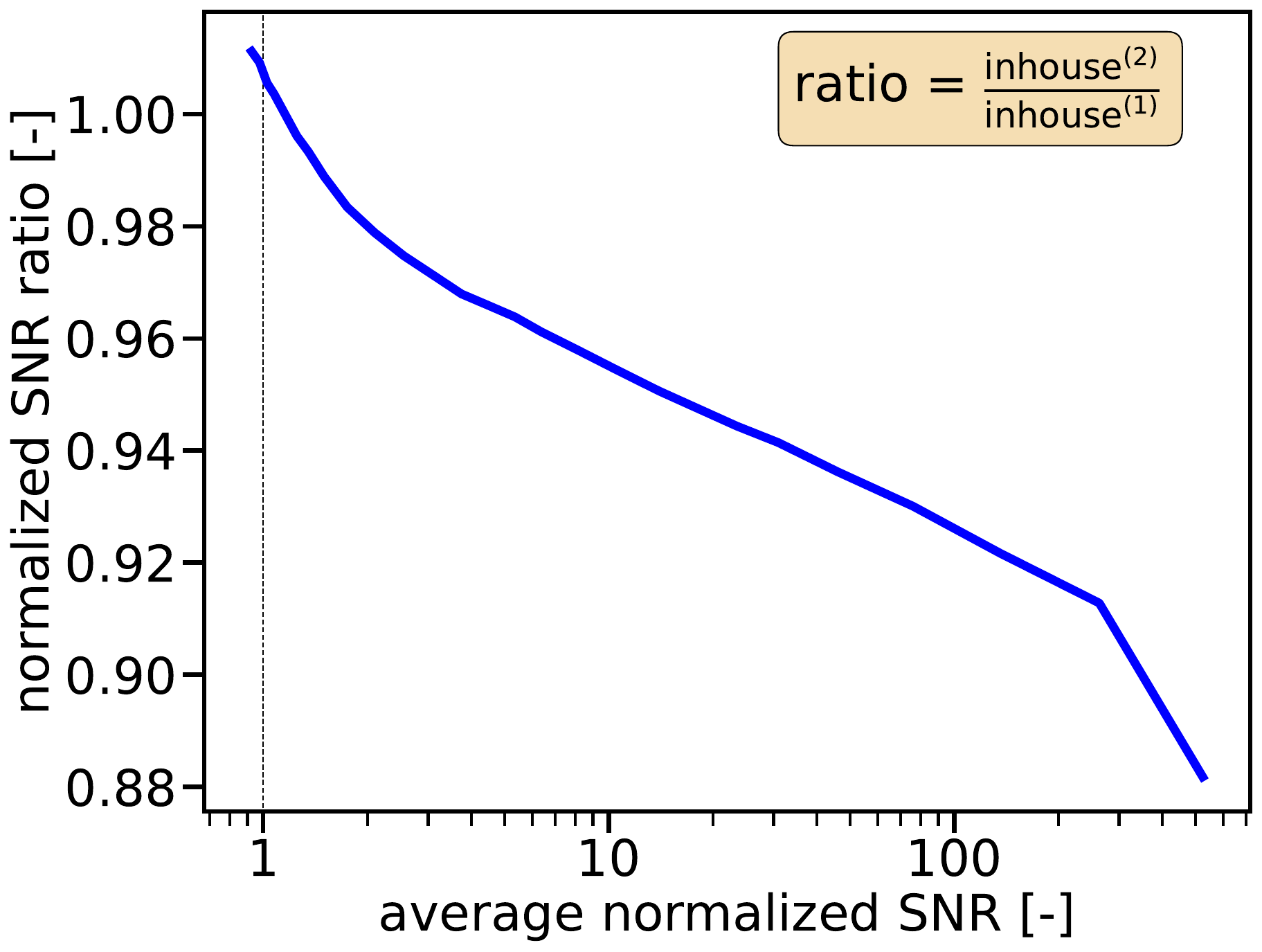}
    \caption{Normalized SNR ratio between inhouse$^{(2)}$ and inhouse$^{(1)}$ as a function of the average normalized SNR for simulations (assuming $\sigma_\mathrm{NB} = 2.5\, N_{\mathrm{p.e.}}/20\,\unit{ns}$).
    All traces with an average normalized SNR exceeding $350$ are grouped into the final interval ($> 350$).
    The $x-$axis is related to the signal strength.
    The normalized SNR ratio approaches unity for weaker signals, indicating similar performance of both in-house algorithms, whereas for larger signals, inhouse$^{(1)}$ yields higher normalized SNR than inhouse$^{(2)}$.
    See the text for further details.}
    \label{fig:ma_vs_fir}
\end{figure}

While this result is expected due to the poorer frequency response of the boxcar window, this difference does not manifest in the signal detection performance, neither in simulations nor for UHECR events.
To gain further insight, the normalized SNR ratio as a function of the average normalized SNR between inhouse$^{(2)}$ and inhouse$^{(1)}$ is plotted in Fig. \ref{fig:ma_vs_fir}.
On the $x-$axis, lower values correspond to weaker signals near the detection threshold, while higher values represent large signals that are easily detected by both algorithms.
As the figure shows, for the weakest signals, the normalized SNR ratio is slightly above unity, and it decreases steadily with increasing signal strength.
This behavior explains why the inhouse$^{(2)}$ algorithm performs comparably to inhouse$^{(1)}$:
for weak signals, those near the detection threshold, both algorithms produce similar normalized SNR values, resulting in similar detection capability.
For larger signals, inhouse$^{(1)}$ achieves significantly higher normalized SNRs, but such signals are already easily detected by both algorithms.
Therefore, in terms of detection performance, there is no advantage to use inhouse$^{(1)}$ over inhouse$^{(2)}$ for larger signals. 
For weak signals, the inhouse$^{(2)}$ algorithm exhibits a slight advantage, which explains its marginally better detection rate in simulations (see Fig. \ref{fig:sim_eval_01}).
This trend is consistent across different values of $\sigma_\mathrm{NB}$ and is observed for UHECR events as well.

It is evident that both in-house algorithms can be successfully applied for data acquisition.
However, the use of an FIR filter with a Hamming window does not lead to the detection of more events.
We argue that the MA filter -- and hence the inhouse$^{(2)}$ algorithm -- is more appropriate, despite its less effective suppression of higher frequencies.
Its hardware implementation is much simpler, and since no detection benefit is gained from the more complex FIR filter, this complexity is not justified for data acquisition.
A general FIR filter becomes advantageous only during post-processing of large signals, where the poorer frequency response of the MA filter begins to matter.

\newpage
\subsection{On the use of standard deviation in triggering algorithms}
\label{subsec:disc:std}
The triggering algorithms defined in Eq.\,(\ref{eq:snr_definitions}) can be simplified by omitting the respective denominators and treating them as unity.
This simplification offers an advantage by eliminating the estimation error associated with calculating the standard deviation.
Note that under this simplification, both reference algorithms are reduced to computing only an MA, which replicates the approach of the first-level trigger currently used at the Auger observatory \cite{Auger2010}.
However, two key conditions must be satisfied in order to neglect standard deviation calculation:
\begin{itemize}
    \item A variable threshold must be implemented to maintain a constant trigger rate.
    \item Changes in the NB, specifically in its standard deviation, must remain gradual and slow.
\end{itemize}
If the NB evolves much more slowly than the response time of the variable threshold, then the influence of standard deviation is absorbed into the variable threshold itself.
Rapid variations in the NB would meanwhile result in either an extremely high trigger rate (due to an increase in the standard deviation of the NB) or a significantly reduced trigger rate (due to an decrease in the standard deviation of the NB).
Fortunately, during standard observation, NB variations are indeed gradual (see Appendix~\ref{appendix:nb} for details), which allows for this simplification.
One might consider applying the same simplification to the $\mathrm{trace}_{\mathrm{mean}}(P_{\mathrm{ma}})$ term in Eqs. (\ref{eq:snr1}) and (\ref{eq:snr2}).
However, this is not feasible because the floating baseline varies rapidly even within individual traces, which makes the calculation of $\mathrm{trace}_{\mathrm{mean}}(P_{\mathrm{ma}})$ necessary.

Note that omitting the denominators does not affect the comparison performed between the in-house algorithms and the reference$^{(1)}$ algorithm, since the standard deviation was estimated using the same method across these three algorithms.
However, it does not apply to reference$^{(2)}$, which was included specifically to demonstrate the necessity of using $\sigma(\mathrm{trace})$ instead of $\sigma(\mathrm{trace}_{\mathrm{filt}})$.
It is also worthy to note that calculating standard deviation in Eq.~(\ref{eq:snr_definitions}) slightly disadvantages the in-house algorithms, as the estimation error in the standard deviation also propagates to the $\mathrm{trace}_{\mathrm{mean}}(P_{\mathrm{ma}})$ term.
However, we do not consider this effect relevant, since the in-house algorithms still clearly outperform the reference$^{(1)}$ algorithm.

At Auger, the presented simplification is possible due to the use of a variable threshold to maintain a $100\,\unit{Hz}$ trigger rate per PMT.
On the other hand, TA uses a fixed SNR threshold of six, which requires explicit computation of the standard deviation.
This approach, however, introduces inaccuracy into the SNR due to the estimation error in the standard deviation.
We conjecture that using a variable threshold to achieve a fixed trigger rate is thus a more robust and preferable solution for FAST.
Currently, the inhouse$^{(2)}$ algorithm is implemented in the hardware of the second generation of FAST telescopes \cite{Hamal2024} in its complete form as defined in Eq.~(\ref{eq:snr2}), including the standard deviation calculation.
However, setting the denominator to unity remains a straightforward adjustment that offers flexibility for future improvements.

\subsection{On the use of an internal trigger for a FAST array}
\label{subsec:disc:trigger}
When calculating trigger thresholds in Sect. \ref{subsec:thresholds}, a fixed trigger rate of $1.25\, \unit{Hz}$ was used for each filter window length $w_{\mathrm{filt}}$ per PMT.
With five different values of $w_{\mathrm{filt}}$ and four PMTs, this corresponds to a maximum total trigger rate of $25\,\unit{Hz}$ per FAST telescope.
Over a typical night measurement of $10$ hours, it yields approximately $188 \, \unit{GB}$ of uncompressed data (assuming $7000$ bins per trace).
When stored in the ROOT format \cite{Brun1997}, which includes compression, the size reduces to about $60 \, \unit{GB}$.
Such a volume does not pose a fundamental storage challenge.
Ultimately, the FAST collaboration plans to deploy a large array of multiple FAST telescopes to cover over an area greater than that of the current largest UHECR experiments.
A major step towards this goal is the creation of ``FAST mini-array'' consisting of six telescopes. 
This array is scheduled for completion by the end of 2026~\cite{Hamal2024}.

For a large array of multiple FAST telescopes, the resulting data volume is still manageable, but will require the dataset to be reduced after each night of observation to prevent data storage overflow.
As a possible solution, a stereo trigger can be implemented.
The stereo trigger identifies time coincidences between signals detected by FAST telescopes in different locations.
Only events detected by at least two FAST telescopes at different locations can be stored.
Since stereo detection is essential for reconstruction of event parameters~\cite{Bradfield2025}, any event without a stereo coincidence can be discarded. 
This simple approach will significantly reduce the dataset generated during each night.
Note that due to the large distance between individual FAST telescopes, the stereo trigger cannot be implemented in real-time because of data transfer limitations.
Therefore, it will be applied automatically after each night's data collection.
This implementation offers an additional benefit: since the analysis is no longer constrained by the hardware, more sophisticated software-based analyses and complex filtering procedures can be applied to further improve event selection.

\section{Conclusion}
In this paper, we have introduced a novel triggering algorithm for FAST, which will be essential for the autonomous operation of the soon-to-be deployed FAST field telescopes. 
Specifically, we presented two in-house algorithms, which differ in their filtering strategies: both employ FIR filters, with either a Hamming window or a boxcar window. 
It was demonstrated using Monte Carlo simulations of EAS and data collected by the current FAST telescope at Auger that both algorithms perform comparably and significantly outperform the approaches based on those implemented by the largest UHECR experiments when applied to FAST data. 
Due to the simpler hardware implementation of the FIR filter with a boxcar window, this method was selected as the most suitable for data acquisition in the second generation FAST telescopes. 

In addition, we presented preliminary results of UHECR detection by the FAST telescope deployed in the southern hemisphere.
A total of $268$ UHECR events were found with the inhouse$^{(2)}$ algorithm between July -- October 2022. 
The relationship between the reconstructed UHECR energy and impact parameter was analysed, suggesting that the individual FAST telescope can detect UHECRs with an energy above $60\,\unit{EeV}$ at distances up to $\approx 20\,\unit{km}$.
These results support FAST as a promising candidate for a next-generation cosmic ray experiment, with the potential to make a substantial contribution to understanding UHECRs.

\clearpage
\appendix
\section{Statistical properties of sample mean and standard deviation}
The expected value and the standard deviation of the sample mean and sample standard deviation are derived here, as these relations are crucial for calibrating parameters of the triggering algorithms.

\subsection{Sample mean}
\label{appendix:calib00_mean}
Let $X_1, \ldots, X_n$ be a sample of $n$ independent observations on a random variable $X \sim \mathcal{N}(\mu,\,\sigma^{2})$. The sample mean is defined as:
\begin{equation}
    e = \frac{1}{n} \sum_{i = 1}^{n} X_i.
\end{equation}
Using the linearity of the expectation operator, the expected value of $e$ is:
\begin{equation}
    \mathbb{E}(e) = \mathbb{E}\bigg(\frac{1}{n} \sum_{i = 1}^{n} X_i\bigg) = \mu.
\end{equation}
Since $X_i$ are uncorrelated, the variance of the sample mean equals:
\begin{equation}
    \mathrm{Var}(e) = \mathrm{Var}\bigg(\frac{1}{n} \sum_{i = 1}^{n} X_i\bigg) = \frac{1}{n^2}\sum_{i = 1}^{n} \mathrm{Var}(X_i) = \frac{\sigma^2}{n}.
\end{equation}
Therefore, the standard deviation of the sample mean is given by:
\begin{equation}
    \mathrm{SD}(e) = \frac{\sigma}{\sqrt{n}}.
    \label{eq:sd_e}
\end{equation}
This derivation can be naturally extended to a weighted average with arbitrary coefficients $h_k$, as used in the FIR filter, yielding the general form expressed in Eq. (\ref{eq:sigma}).

\subsection{Sample standard deviation}
\label{appendix:calib00_std}
Let $X_1, \ldots, X_n$ be a sample of $n$ independent observations on a random variable $X \sim \mathcal{N}(\mu,\,\sigma^{2})$. The sample standard deviation is defined as:
\begin{equation}
    s = \sqrt{\frac{1}{n-1} \sum_{i = 1}^{n} (X_i - e)^2 },
\end{equation}
where $e$ is the sample mean. 
It is known that 
\begin{equation}
    s \sim \frac{\sigma}{\sqrt{n-1}}\chi_{n-1},    
\end{equation}
where $\chi_{n-1}$ denotes the chi distribution with $n-1$ degrees of freedom \cite{Cochran1934}. 
The expected value of the chi distribution is given by $\mathbb{E}(\chi_{n-1}) = \sqrt{2} \frac{\Gamma(n/2)} {\Gamma((n-1)/2)}$ \cite{Forbes2011}, where $\Gamma$ is the gamma function.
Substituting this into the distribution of~$s$, the expected value of the sample standard deviation becomes:
\begin{equation}
    \label{eq:es}
    \mathbb{E}(s) = \sigma \sqrt{\frac{2}{n-1}} \frac{\Gamma(\frac{n}{2})}{\Gamma(\frac{n-1}{2})}.
\end{equation}
To determine the variability of $s$, we use the basic property of the variance $\mathrm{Var}(s) = \mathbb{E}(s^2) - \mathbb{E}(s)^2$.
Since $\mathbb{E}(s^2) = \sigma^2$, it follows:
\begin{equation}
    \mathrm{Var}(s) = \sigma^2 - \Bigg( \sigma \sqrt{\frac{2}{n-1}} \frac{\Gamma(\frac{n}{2})}{\Gamma(\frac{n-1}{2})} \Bigg)^2, 
\end{equation}
and the standard deviation of the sample standard deviation is thus:
\begin{equation}
    \mathrm{SD}(s) = \sigma \sqrt{1 - \frac{2}{n-1} \Bigg( \frac{\Gamma(\frac{n}{2})}{\Gamma(\frac{n-1}{2})} \Bigg)^2 }.
\end{equation}
The relative errors of the expected value and standard deviation of the sample standard deviation $s$ are then given by:
\begin{equation}
    \label{eq:rel_error}
    \frac{\sigma - \mathbb{E}(s)}{\sigma} = 1 - \sqrt{\frac{2}{n-1}} \frac{\Gamma(\frac{n}{2})}{\Gamma(\frac{n-1}{2})} \approx \frac{1}{4n}, \quad
    \frac{\mathrm{SD}(s)}{\sigma} = \sqrt{1 - \frac{2}{n-1} \Bigg( \frac{\Gamma(\frac{n}{2})}{\Gamma(\frac{n-1}{2})} \Bigg)^2 } \approx \frac{1}{\sqrt{2n}},
\end{equation}
where the approximations hold for large $n$, based on asymptotic expansions of the gamma function \cite{Abramowitz1965}. 
Specifically, the approximation of the gamma function ratio used here corresponds to Eq. (6.1.47) in \cite{Abramowitz1965}.
Note that the non-zero relative error of the expected value confirms that the sample standard deviation $s$ is a biased estimator~of~$\sigma$.

\section{Parameter calibration of triggering algorithms}
\label{appendix:calib}

\subsection{Parameter \texorpdfstring{$w_{\mathrm{std}}$}{w}}
\label{appendix:calib01}
When calculating the standard deviation, a sufficiently large amount of data is required to ensure an accurate estimate. 
The length of the window $w_{\mathrm{std}}$ (see the scheme in Fig. \ref{fig:alg_scheme}) is therefore determined by the acceptable level of error in the standard deviation estimate.
To quantify this, we use the approximation from Eq. (\ref{eq:rel_error}).
For a window size of $n := w_{\mathrm{std}} = 2048$, this yields $\mathrm{SD}(s)/\sigma \approx 0.016$.
In other words, once in every $100$ million calculations, the sample standard deviation may deviate from the true value by approximately~$9\%$.
This indicates that using $2048$ bins provides a sufficiently accurate estimate of $\sigma$.
The chosen value of $w_{\mathrm{std}}$ thus represents a practical compromise between accuracy and computational or data acquisition constraints.

\subsection{Parameters \texorpdfstring{$w_{\mathrm{nan}}$}{w} and \texorpdfstring{$w_{\mathrm{ma}}$}{w}}
\label{appendix:calib02}

The calculation of the baseline value $\mathrm{trace}_{\mathrm{mean}}(P_{\mathrm{ma}})$ in Eq. (\ref{eq:snr}), used for SNR calculation, depends on the parameters $w_{\mathrm{nan}}$ and~$w_{\mathrm{ma}}$, as the baseline is obtained by averaging trace values over the interval $[P_{\mathrm{nan}} - w_{\mathrm{ma}}, P_{\mathrm{nan}}]$.
Here, $w_{\mathrm{nan}}$ determines the position~$P_{\mathrm{nan}}$, which serves as the endpoint of the averaging window, while $w_{\mathrm{ma}}$ specifies the number of samples used in average.
To calibrate these parameters, our goal is to minimize the error introduced by the baseline estimate.

To this end, we decompose the trace into two independent components: a floating baseline without noise (see Fig. \ref{fig:baselines}), and the NB.
This decomposition is consistent with the trace model used throughout the paper.
The estimated baseline value $\mathrm{trace}_{\mathrm{mean}}(P_{\mathrm{ma}})$ can be then expressed as:
\begin{equation}
    \mathrm{trace}_{\mathrm{mean}}(P_{\mathrm{ma}}) = \mathrm{baseline}(P_{\mathrm{test}}) - (\mathrm{error_{baseline}} + \mathrm{error}_{\mathrm{NB}}),
\end{equation}
where $\mathrm{baseline}(P_{\mathrm{test}})$ denotes the true baseline value at the signal position $P_{\mathrm{test}}$, $\mathrm{error_{baseline}}$ is the estimation error of the baseline value (excluding noise), and $\mathrm{error}_{\mathrm{NB}}$ is the error due to averaging the NB.
Therefore, the total estimation error is decomposed into two independent contributions.
Note that in practice, $\mathrm{baseline}(P_{\mathrm{test}})$ cannot be directly computed because it overlaps with the potential signal.

As a result, the SNR calculated by the in-house algorithms becomes:
\begin{equation}
    \mathrm{SNR} = \frac{  \mathrm{trace}_{\mathrm{filt}} (P_{\mathrm{test}})  - \mathrm{trace}_{\mathrm{mean}}(P_{\mathrm{ma}}) }
    {\sigma(\mathrm{trace})\sqrt{\sum_{k = 0}^{m-1} h_k^2}} = 
    \frac{  \mathrm{trace}_{\mathrm{filt}} (P_{\mathrm{test}})  - \mathrm{baseline}(P_{\mathrm{test}}) }
    {\sigma(\mathrm{trace})\sqrt{\sum_{k = 0}^{m-1} h_k^2}} 
    + \frac{ \mathrm{error_{baseline} + \mathrm{error}_{\mathrm{NB}}} }
    {\sigma(\mathrm{trace})\sqrt{\sum_{k = 0}^{m-1} h_k^2}}. 
    \label{eq:snr_error}
\end{equation}
Our aim is to minimize the final term in Eq. (\ref{eq:snr_error}); specifically, we search for values of $w_{\mathrm{nan}}$ and $w_{\mathrm{ma}}$ that minimize its standard deviation.
Since the filter coefficients $h_k$ are fixed for a given FIR filter, their sum is constant and can be set to unity for optimization purposes.
Given that $\mathrm{error_{baseline}}$ and $\mathrm{error}_{\mathrm{NB}}$ are independent, their combined standard deviation is obtained as a quadratic sum of their individual contributions.
Therefore, minimizing the final term in Eq. (\ref{eq:snr_error}) is equivalent to minimizing the total standard deviation $\sigma_{\mathrm{total\,error}}$:
\begin{equation}
    \sigma_{\mathrm{total\,error}} = 
    \frac{\sqrt{\sigma^2(\mathrm{error_{baseline}} ) + \sigma^2(\mathrm{error_{NB}} ) }}
    {\sigma_{\mathrm{NB}}},
    \label{eq:std_total}
\end{equation}
where we assume no error in $\sigma(\mathrm{trace})$ calculation, i.e., $\sigma(\mathrm{trace}) = \sigma_{\mathrm{NB}}$.
This assumption does not affect the optimization process because $\sigma(\mathrm{trace})$ is independent of $w_{\mathrm{nan}}$ and $w_{\mathrm{ma}}$.

While the standard deviation of the error introduced by averaging the NB can be calculated analytically (see Eq.~(\ref{eq:sd_e})):
\begin{equation}
    \sigma(\mathrm{error_{NB}}) = \frac{\sigma_{\mathrm{NB}}}{\sqrt{w_{\mathrm{ma}}}},
\end{equation}
the value of $\sigma(\mathrm{error_{baseline}})$ must be computed numerically.
For each baseline shown in Fig. \ref{fig:baselines}, we evaluate $\mathrm{error_{baseline}}$, assuming an arbitrary but fixed values of $w_{\mathrm{nan}}$ and $w_{\mathrm{ma}}$.
It is defined as the difference between the true baseline value at $P_{\mathrm{test}}$ and the average of baseline values over the interval $[P_{\mathrm{nan}} - w_{\mathrm{ma}}, P_{\mathrm{nan}}]$.
This procedure is applied for every bin in the trace, excluding edge regions where averaging is not possible due to insufficient data (see the scheme in Fig.~\ref{fig:alg_scheme}).
In total, this yields over a billion values of $\mathrm{error_{baseline}}$, from which the standard deviation $\sigma(\mathrm{error_{baseline}})$ is computed.

\begin{figure}[htb!]
    \centering
    \includegraphics[width=0.82\linewidth]{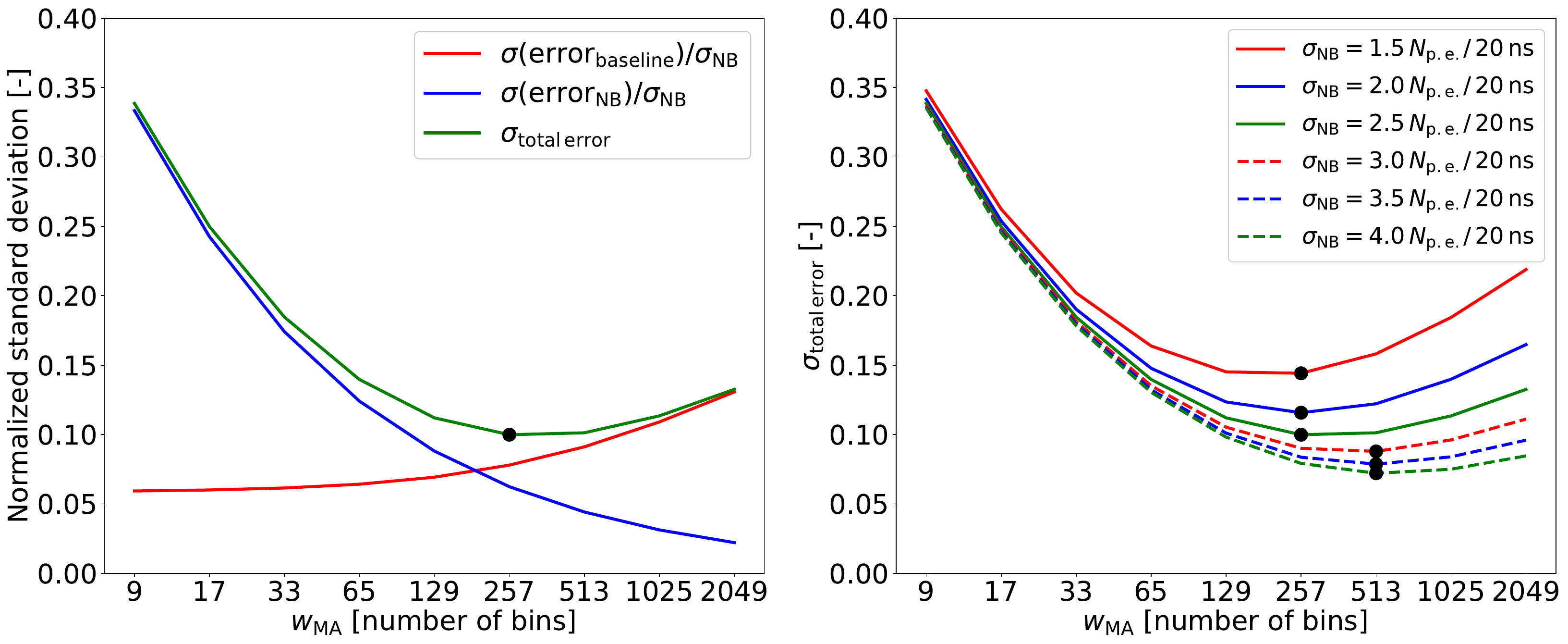}
    \caption{Total standard deviation $\sigma_{\mathrm{total\,error}}$ as a function of the parameter $w_{\mathrm{ma}}$.
    Left panel: Normalized standard deviations of the individual contributions $\sigma(\mathrm{error_{baseline}}) / \sigma_{\mathrm{NB}}$ and $\sigma(\mathrm{error_{NB}})/ \sigma_{\mathrm{NB}}$, along with the total error $\sigma_{\mathrm{total\,error}}$, evaluated for $w_{\mathrm{nan}} = 256$ and $\sigma_{\mathrm{NB}} = 2.5 \, N_{\mathrm{p.e.}}/20\,\unit{ns}$.
    The minimum, corresponding to $w_{\mathrm{ma}} = 257$, is indicated by a black circle.
    Right panel: Total error $\sigma_{\mathrm{total\,error}}$ for $w_{\mathrm{nan}} = 256$ across various noise levels $\sigma_{\mathrm{NB}}$.
    Optimal values of $w_{\mathrm{ma}}$ for each $\sigma_{\mathrm{NB}}$ are again marked by black circles.}
    \label{fig:app:wma}
\end{figure}

The left panel of Fig. \ref{fig:app:wma} shows the normalized standard deviations $\sigma(\mathrm{error_{baseline}}) / \sigma_{\mathrm{NB}}$, $\sigma(\mathrm{error_{NB}})/ \sigma_{\mathrm{NB}}$, and $\sigma_{\mathrm{total\,error}}$ for $w_{\mathrm{nan}} = 256$ and $\sigma_{\mathrm{NB}} = 2.5 \, N_{\mathrm{p.e.}}/20\,\unit{ns}$ across various values of $w_{\mathrm{ma}} = 2^k + 1, k = 3, \dots, 11$.
The first two normalized quantities represent the individual contributions of $\mathrm{error_{baseline}}$ and $\mathrm{error_{NB}}$, respectively.
As $w_{\mathrm{ma}}$ increases, $\mathrm{error_{baseline}}$ also increases because the baseline is estimated using points that are further away from the test position $P_{\mathrm{test}}$.
Moreover, the broader averaging window smooths the baseline more, further increasing the error.
On the contrary, $\mathrm{error_{NB}}$ decreases with increasing $w_{\mathrm{ma}}$, as the larger number of samples used in the mean calculation reduces statistical fluctuations of the noise.

These opposing trends yield an optimal value of $w_{\mathrm{ma}} = 257$ (marked by a black circle), where $\sigma_{\mathrm{total\,error}}$ reaches its minimum.
However, the difference compared to $w_{\mathrm{ma}} = 513$ is negligible.
This behavior is consistent across various values of $\sigma_{\mathrm{NB}}$, as demonstrated in the right panel of Fig. \ref{fig:app:wma}. 
Optimal values are again marked with black circles and are either $257$ or $513$ bins.
Note that lower $\sigma_{\mathrm{NB}}$ leads to a larger $\sigma_{\mathrm{total\,error}}$.
This is expected as a lower NB level makes the floating baseline more pronounced, thus increasing the baseline estimation error. 
This effect explains the rise in threshold values at low $\sigma_{\mathrm{NB}}$, as observed in Fig.~\ref{fig:thresholds} and Fig.~\ref{fig:app:thresholds}.

To objectively determine the optimal $w_{\mathrm{ma}}$, 
we compute a weighted sum of $\sigma_{\mathrm{total\,error}}$ over all considered values of $\sigma_{\mathrm{NB}}$.
Weights are based on the distribution of the NB standard deviation in data collected from 4.7.2022 to 24.10.2022.
The resulting normalized weights are listed in Table \ref{tab:std}.
The same time period is used for the UHECR event analysis in Sect. \ref{subsec:uhecr}.

\begin{table}[h!]
    \caption{Normalized weights of the noise background standard deviation calculated using acquired data over the period from 4.7.2022 to 24.10.2022.}
    \centering
    \begin{tabular}{|l||c|c|c|c|c|c|}
    \hline
    $\sigma_{\mathrm{NB}}$ [$N_{\mathrm{p.e.}}/20\,\unit{ns}$] & 1.5 & 2.0 & 2.5 & 3.0 & 3.5 & 4.0 \\
    \hline
    normalized weight & 0.0798 & 0.4718 & 0.2906 & 0.1207 & 0.0258 & 0.0113 \\
    \hline
    \end{tabular}
    \label{tab:std}
\end{table}

\begin{figure}[htb!]
    \centering
    \includegraphics[width=0.41\linewidth]{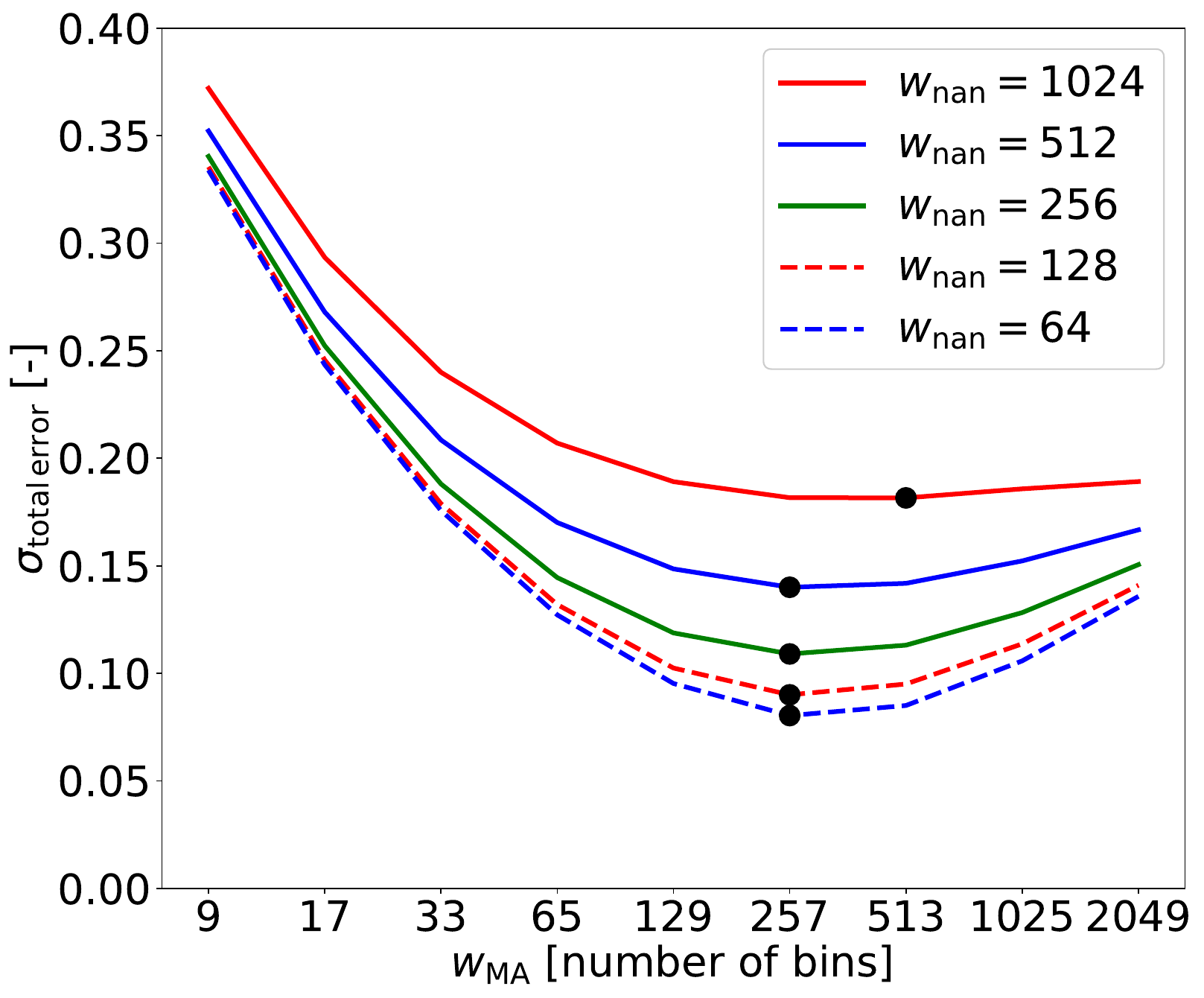}
    \caption{Total standard deviation $\sigma_{\mathrm{total\,error}}$ as a function of the parameter $w_{\mathrm{ma}}$ for five different values of $w_{\mathrm{nan}} = 64, 128, 256, 512$, and $1024$.
    A weighted sum over different $\sigma_{\mathrm{NB}}$ values is applied using the weights from Table \ref{tab:std}.
    Optimal values of $w_{\mathrm{MA}}$ for each $w_{\mathrm{nan}}$, where $\sigma_{\mathrm{total\,error}}$ reaches its minimum, are indicated by black circles.}
    \label{fig:app:wnan_wma}
\end{figure}

Figure \ref{fig:app:wnan_wma} shows the minimization of $\sigma_{\mathrm{total\,error}}$ as a function of $w_{\mathrm{ma}}$, for $w_{\mathrm{nan}} = 64, 128, 256, 512$, and $1024$. 
A~weighted sum across different $\sigma_{\mathrm{NB}}$ values is applied.
The results are consistent with those presented in Fig.~\ref{fig:app:wma}. 
Optimal values of $w_{\mathrm{ma}}$ are indicated by black circles, with $w_{\mathrm{ma}} = 257$ being the best choice across the majority of tested $w_{\mathrm{nan}}$ values, although the differences between $w_{\mathrm{ma}} = 513$ are negligible.
Moreover, increasing $w_{\mathrm{nan}}$ leads to higher $\sigma_{\mathrm{total\,error}}$.
This effect is analogous to increasing $w_{\mathrm{ma}}$, as the baseline is estimated using points further from the test position $P_{\mathrm{test}}$.

Based on these findings, we specify the parameters $w_{\mathrm{nan}}$ and $w_{\mathrm{ma}}$.
If $w_{\mathrm{nan}}$ is too small, the averaging window may include part of the signal, leading to a significant reduction in the SNR.
If $w_{\mathrm{nan}}$ is too large, it introduces an increased baseline estimation error, again resulting in an inaccurate SNR calculation.
To balance these effects, we select $w_{\mathrm{nan}} = 256$, which is far enough from the potential signal, while still maintaining acceptable total error quantified by $\sigma_{\mathrm{total\,error}}$.

For $w_{\mathrm{ma}}$, the optimization indicates $257$ bins as optimal.
However, given the minimal difference in $\sigma_{\mathrm{total\,error}}$, when using $w_{\mathrm{ma}} = 513$, we select the larger value.
This choice further reduces the effect of potential signal contamination in the baseline calculation $\mathrm{trace}_{\mathrm{mean}}(P_{\mathrm{ma}})$, since the averaging includes more distant points from the signal location.

\subsection{Frequency response of FIR filters}
\label{appendix:calib03}
\begin{figure}[htb!]
    \centering
    \includegraphics[width=0.80\linewidth]{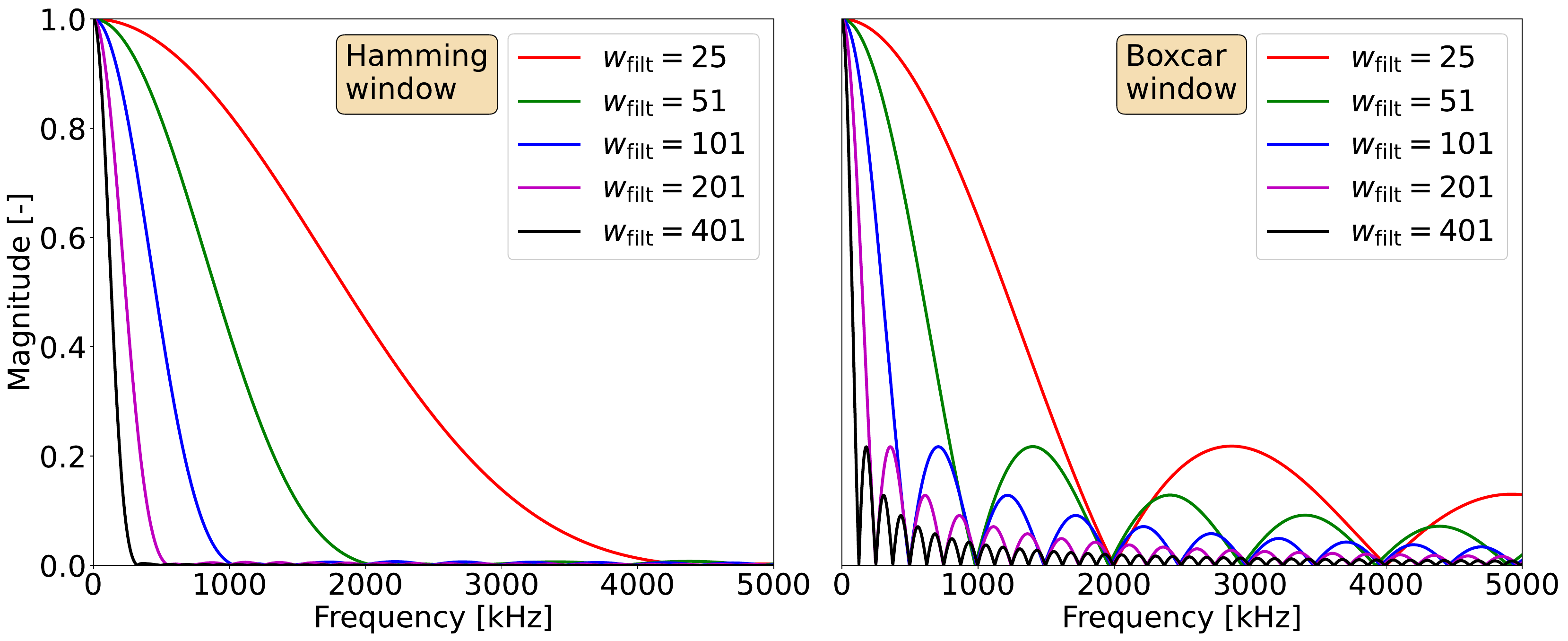}
    \caption{Frequency response of FIR filters. 
    The left panel shows the response for the Hamming window with a cutoff frequency of $100\,\unit{kHz}$, while the right panel corresponds to the boxcar window, i.e., the MA filter. 
    The Hamming window demonstrates better suppression of higher frequencies.}
    \label{fig:app:freq}
\end{figure}

Figure \ref{fig:app:freq} shows the frequency response of both FIR filters used in this study: one with a Hamming window and the other with a boxcar window, for five different filter lengths $w_{\mathrm{filt}} = 25, 51, 101, 201$, and $401$ bins.
Note that the boxcar window corresponds to an MA filter.
The FIR filter with a Hamming window is designed with a cutoff frequency of $100\, \unit{kHz}$ for all values of $w_{\mathrm{filt}}$. 
The cutoff frequency was chosen to be sufficiently low such that further reduction has no significant effect on the filtered frequencies, regardless of $w_{\mathrm{filt}}$.
As a result, the frequency attenuation is primarily determined by the filter lengths; specifically, longer filter lengths are used to suppress lower frequencies, which effectively removes the noise background.

It is evident from the figure that the MA filter provides poorer high-frequency attenuation compared to the FIR filter with a Hamming window.
While the Hamming window achieves almost complete suppression of high frequencies, the MA exhibits significant ripples in the frequency response, a well-known artifact of the boxcar window.

The frequency responses in Fig. \ref{fig:app:freq} confirm that the selected combination of filter lengths $w_{\mathrm{filt}}$ and cutoff frequency (for the Hamming window) is appropriate for the requirements of data acquisition. 
Different frequencies are suppressed by different window lengths, which allows effective detection of both short and long signals.

\section{Complementary analysis and results}

\subsection{Standard deviation of the filtered trace}
\label{appendix:std}

To analyse the variability of $\sigma(\mathrm{trace_{\mathrm{filt}}})$ and compare it with that of $\sigma(\mathrm{trace}) \big(\sum_{k = 0}^{m-1} h_k^2\big)^{\frac{1}{2}}$, the ideal approach would be to compute both quantities analytically.
While the variability derivation of $\sigma(\mathrm{trace}) \big(\sum_{k = 0}^{m-1} h_k^2\big)^{\frac{1}{2}}$ is relatively straightforward (see Eq. (\ref{eq:rel_error}) in Appendix \ref{appendix:calib00_std}), we were unable to obtain an analytical solution for the variability of $\sigma(\mathrm{trace_{\mathrm{filt}}})$. 
Therefore, a numerical analysis is provided here for both quantities.

One million traces are generated from a normal distribution $\mathcal{N}(0,\,\sigma^{2})$ with $\sigma = 2.5 \, N_{\mathrm{p.e.}}/20\,\unit{ns}$.
Each trace is filtered using an MA filter with window length $w_{\mathrm{filt}} = 101$.
For each filtered trace, the standard deviation is computed over a window of $w_{\mathrm{std}} = 2048$ bins, being consistent with the parameters used in the triggering algorithms.
This yields a distribution of one million $\sigma(\mathrm{trace_{\mathrm{filt}}})$ values.
The same procedure is applied to the original (unfiltered) traces to obtain one million values of $\sigma(\mathrm{trace})/ \sqrt{w_{\mathrm{filt}}}$, noting that for an MA filter, we can write $\big(\sum_{k = 0}^{m-1} h_k^2\big)^{\frac{1}{2}} = 1/ \sqrt{w_{\mathrm{filt}}}$.

\begin{figure}[ht!]
    \centering
    \includegraphics[width=0.40\linewidth]{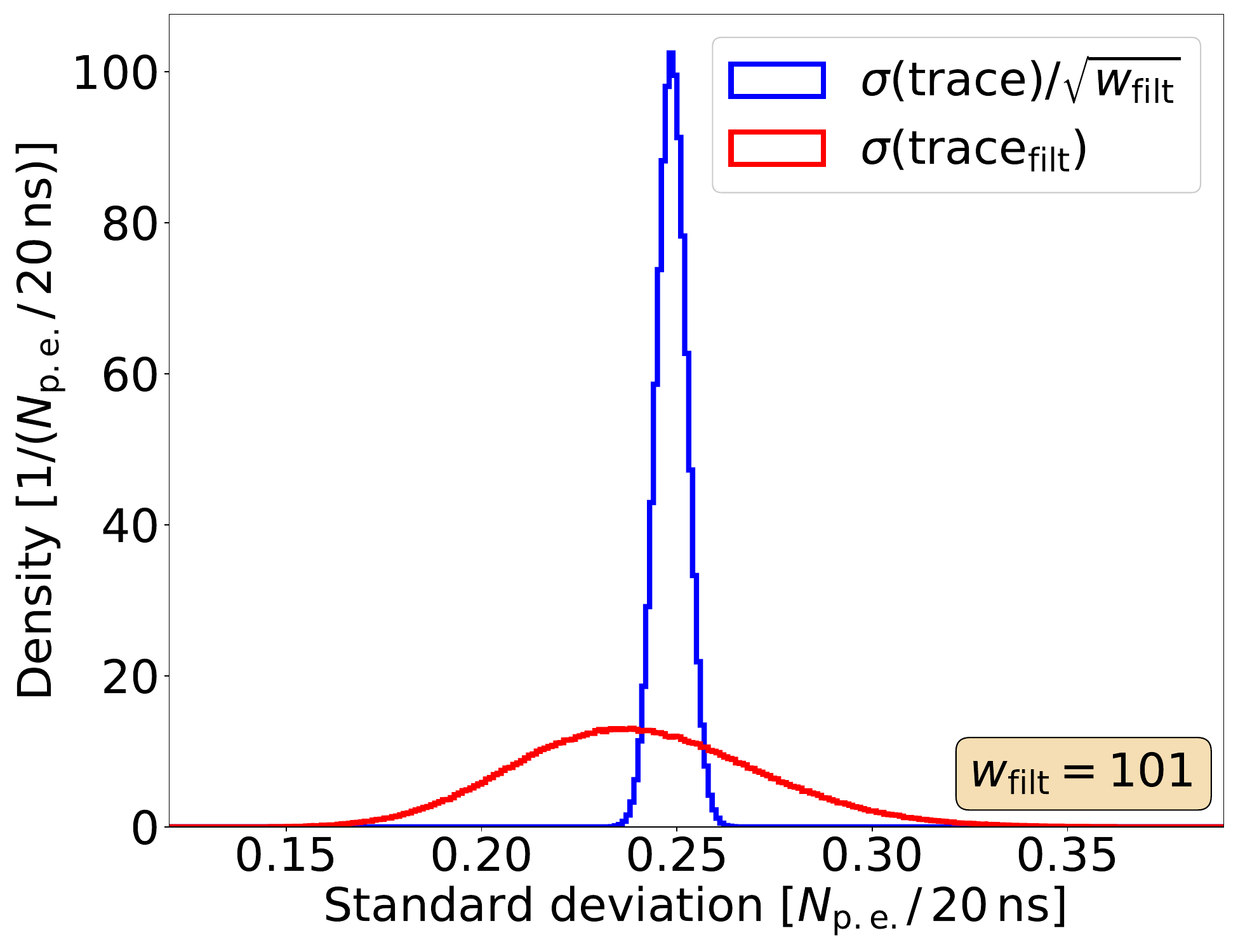}
    \caption{PDFs of $\sigma(\mathrm{trace_{\mathrm{filt}}})$ and $\sigma(\mathrm{trace}) / \sqrt{w_{\mathrm{filt}}}$ for $w_{\mathrm{filt}} = 101$. 
    Standard deviations are computed over $w_{\mathrm{std}} = 2048$ bins. 
    The filtered trace exhibits substantially higher variability compared to the predicted standard deviation from the unfiltered trace.}
    \label{fig:app:std}
\end{figure}

Figure \ref{fig:app:std} shows calculated PDFs of $\sigma(\mathrm{trace_{\mathrm{filt}}})$ and $\sigma(\mathrm{trace}) / \sqrt{w_{\mathrm{filt}}}$, both from one million values.
The distribution of $\sigma(\mathrm{trace_{\mathrm{filt}}})$ is significantly broader, which demonstrates its much higher variability.
Specifically, the standard deviation of $\sigma(\mathrm{trace}) / \sqrt{w_{\mathrm{filt}}}$ is approximately 
$0.0039 \, N_{\mathrm{p.e.}}/20\,\unit{ns}$, while for $\sigma(\mathrm{trace_{\mathrm{filt}}})$, it is approximately $0.0309\, N_{\mathrm{p.e.}}/20\,\unit{ns}$ -- an order of magnitude larger.
Extremely low values of $\sigma(\mathrm{trace_{\mathrm{filt}}})$ leads to artificially high SNR values for the noise background (see Eq.~(\ref{eq:snr4})), therefore negatively affecting the performance of the reference$^{(2)}$ algorithm.
These results are consistent regardless of different types of FIR filters and filter lengths~$w_{\mathrm{filt}}$.

\newpage
\subsection{Threshold values of triggering algorithms}
\label{appendix:thrsh_calc}
\begin{figure}[ht!]
    \centering
    \includegraphics[width=0.65\linewidth]{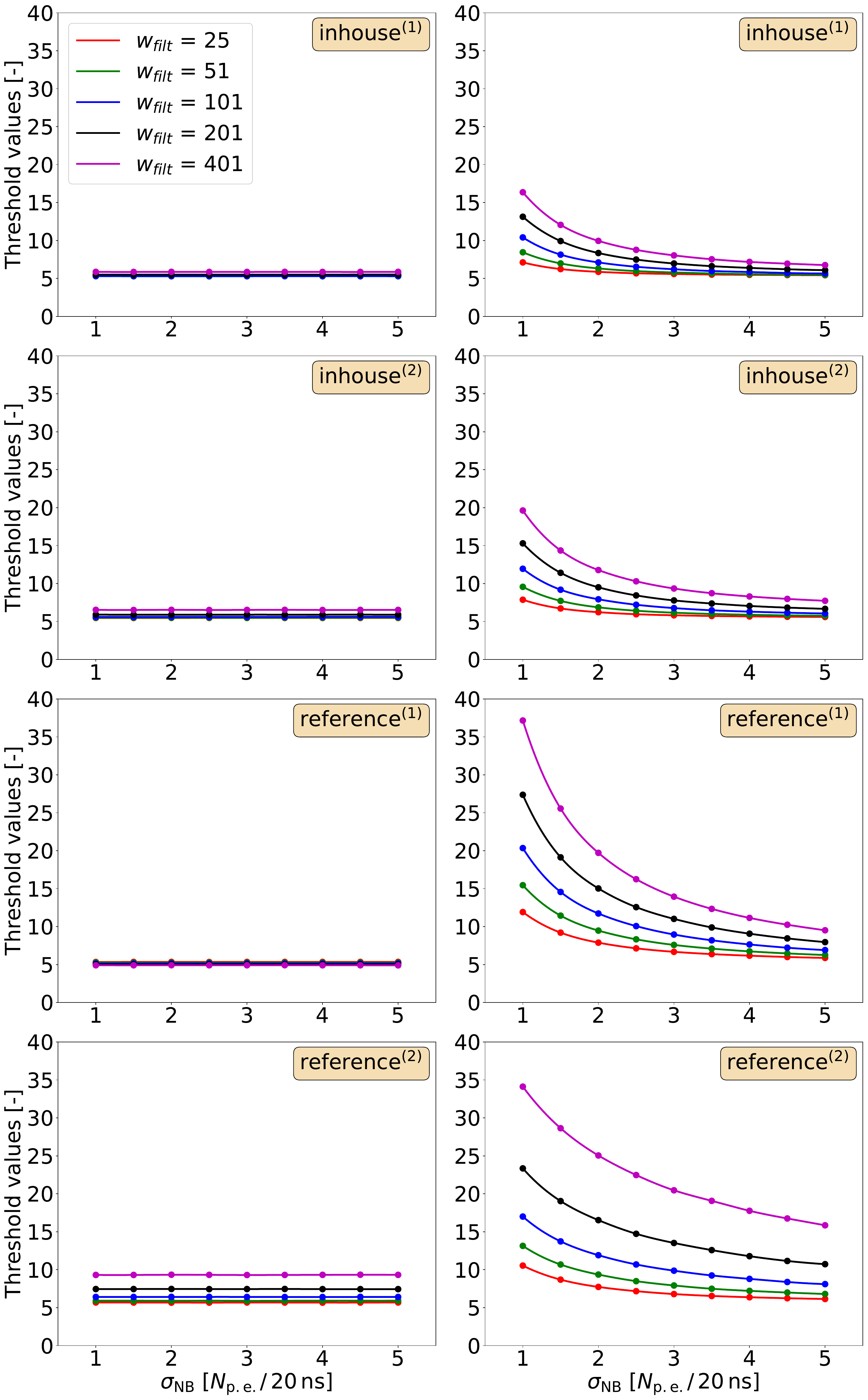}
    \caption{Threshold values required to maintain a trigger rate of $1.25\,\unit{Hz}$ per a single PMT and filter length $w_{\mathrm{filt}}$ for all triggering algorithms, evaluated across different values of $\sigma_{\mathrm{NB}}$ and $w_{\mathrm{filt}}$. 
    Left panels: Thresholds obtained using an NB without floating baselines. 
    Right panels: Thresholds obtained using an NB with floating baselines, representing realistic conditions during FAST data acquisition.}
    \label{fig:app:thresholds}
\end{figure}

Figure \ref{fig:app:thresholds} shows the threshold values required to maintain a trigger rate of $1.25\,\unit{Hz}$ per a single PMT and filter length $w_{\mathrm{filt}}$, for all triggering algorithms.
Thresholds are calculated across various noise background levels $\sigma_{\mathrm{NB}}$ and filter lengths $w_{\mathrm{filt}}$.
The left panels of Fig. \ref{fig:app:thresholds} present the threshold values obtained when the NB is generated without floating baselines.
In this case, the thresholds remain nearly constant.
In contrast, the right panels include realistic floating baselines, which cause the threshold values to vary significantly with both $\sigma_{\mathrm{NB}}$ and $w_{\mathrm{filt}}$.

\newpage
\subsection{Probability density function of \texorpdfstring{$X_{\mathrm{max}}$}{Xmax}}
\label{appendix:pdfs}

\begin{figure}[ht!]
    \centering
    \includegraphics[width=0.42\linewidth]{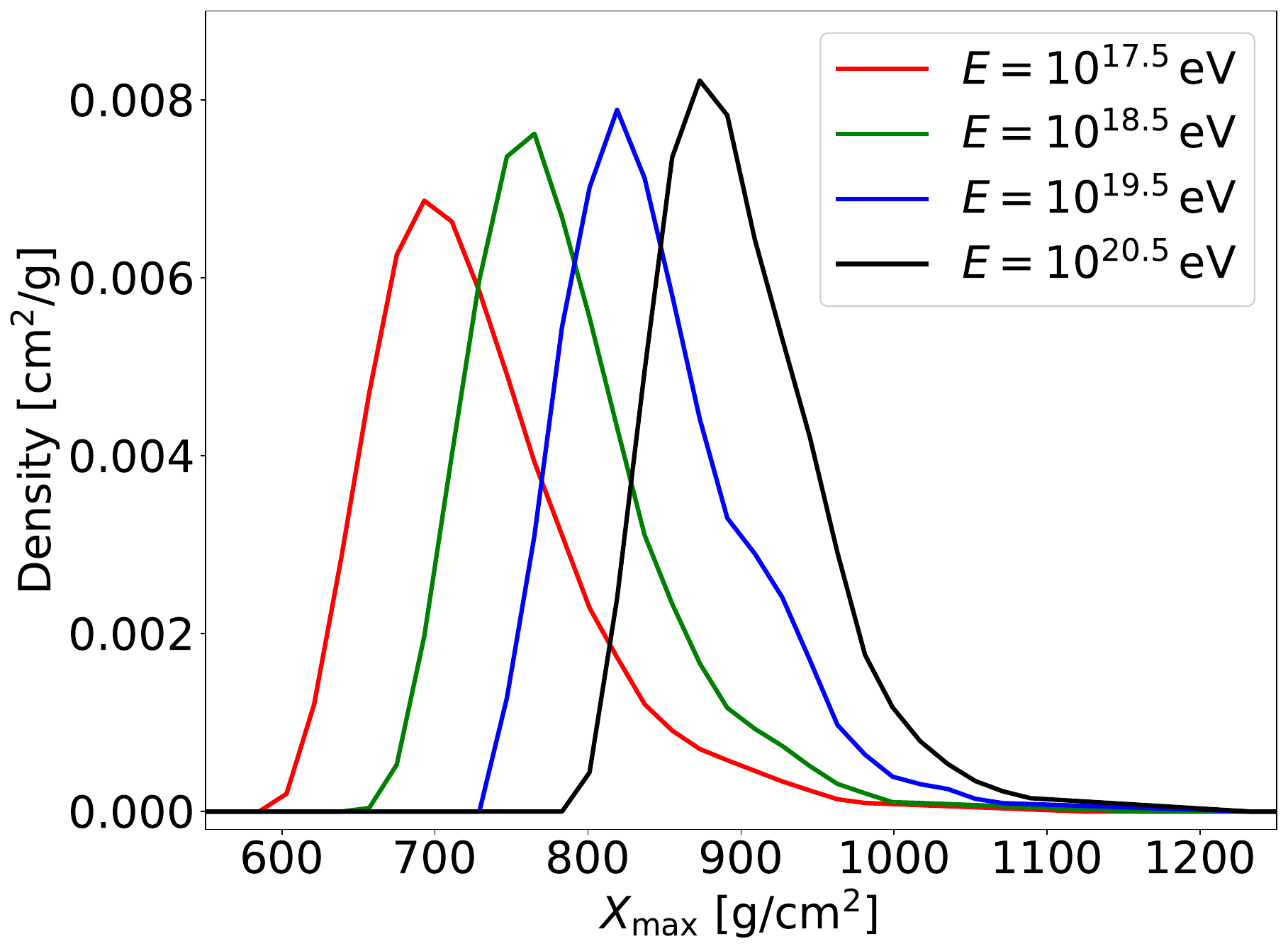}
    \caption{PDFs of $X_{\mathrm{max}}$ obtained from CONEX simulations for protons as primary particles at selected energies. 
    For energies in the interval [$10^{17.5}, 10^{20.5}]\,\unit{eV}$, PDFs are interpolated using a linear fit in $\log_{10}(E)$.}
    \label{fig:app:pdfs}
\end{figure}

The PDFs of the $X_{\mathrm{max}}$ parameter for various energies are shown in Fig. \ref{fig:app:pdfs}.
These PDFs are derived from CONEX simulations \cite{Bergmann2007} assuming protons as primary particles.
Simulations were performed for four energies: $E = 10^{17.5}, 10^{18.5}, 10^{19.5}$, and $10^{20.5}\, \unit{eV}$.
PDFs for intermediate energies are obtained by linear interpolation in $\log_{10}(E)$ scale.
For instance, the PDF for $E = 10^{18}\, \unit{eV}$ is calculated as $0.5$ times the PDF at $E = 10^{17.5}\, \unit{eV}$ plus $0.5$ times the PDF at $E = 10^{18.5}\, \unit{eV}$. 
Therefore, a continuous model of the $X_{\mathrm{max}}$ PDF is available across the entire energy range $[10^{17.5}, 10^{20.5}] \, \unit{eV}$, allowing for the generation of $X_{\mathrm{max}}$ values corresponding to any randomly sampled energy within this interval.

\subsection{Detection ratio of simulations without floating baselines}
\label{appendix:detection_ratio}
\begin{figure}[ht!]
    \centering
    \includegraphics[width=0.40\linewidth]{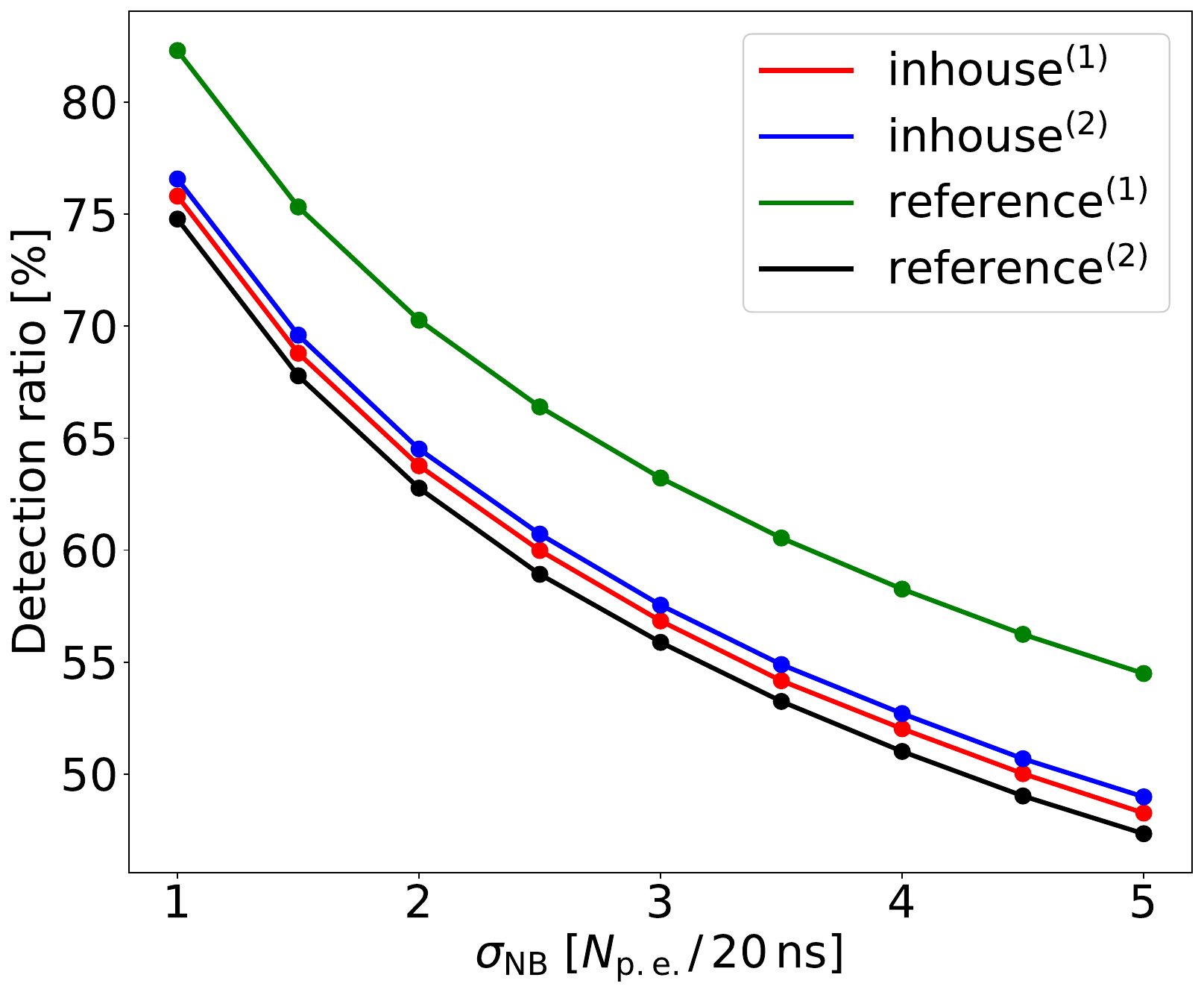}
    \caption{Detection ratio of UHECR simulations for various triggering algorithms as a function of $\sigma_{\mathrm{NB}}$.
    Threshold values are obtained using a noise background without floating baselines. 
    Under these idealized conditions, the reference$^{(1)}$ algorithm consistently outperforms the in-house algorithms regardless of the noise background level $\sigma_{\mathrm{NB}}$.}
    \label{fig:app:detection_ratio}
\end{figure}

Figure \ref{fig:app:detection_ratio} shows the detection ratio across all triggering algorithms when floating baselines are not included in the analysis, neither in the threshold calculations nor in the UHECR simulations.
This setup is equivalent to having a sufficiently high signal-to-noise ratio, as achieved by larger telescopes such as Auger or TA.
As expected, the reference$^{(1)}$ algorithm outperforms the in-house algorithms.
This behavior arises from the estimation of $\mathrm{trace}_{\mathrm{mean}}(P_{\mathrm{ma}})$ value in Eqs. (\ref{eq:snr1}) and (\ref{eq:snr2}), which should be zero when the floating baseline is absent.
In this case, the in-house algorithms introduce an unnecessary error in estimating this zero baseline level, which reduces their effectiveness relative to the reference$^{(1)}$ algorithm.

\subsection{Noise background}
\label{appendix:nb}

Here, we analyse whether the standard deviation of the NB evolves slowly enough to be treated as constant during nighttime observation.
If this condition holds, the denominators in Eq. (\ref{eq:snr_definitions}) can be omitted, simplifying all four triggering algorithms.
To evaluate this, the variation of the NB is examined during a typical night of data acquisition.
For each recorded event, four values of the standard deviation $\sigma_{\mathrm{event}}$ are calculated, one for each PMT, over the first $1024$ bins.

\begin{figure}[ht!]
    \centering
    \includegraphics[width=0.80\linewidth]{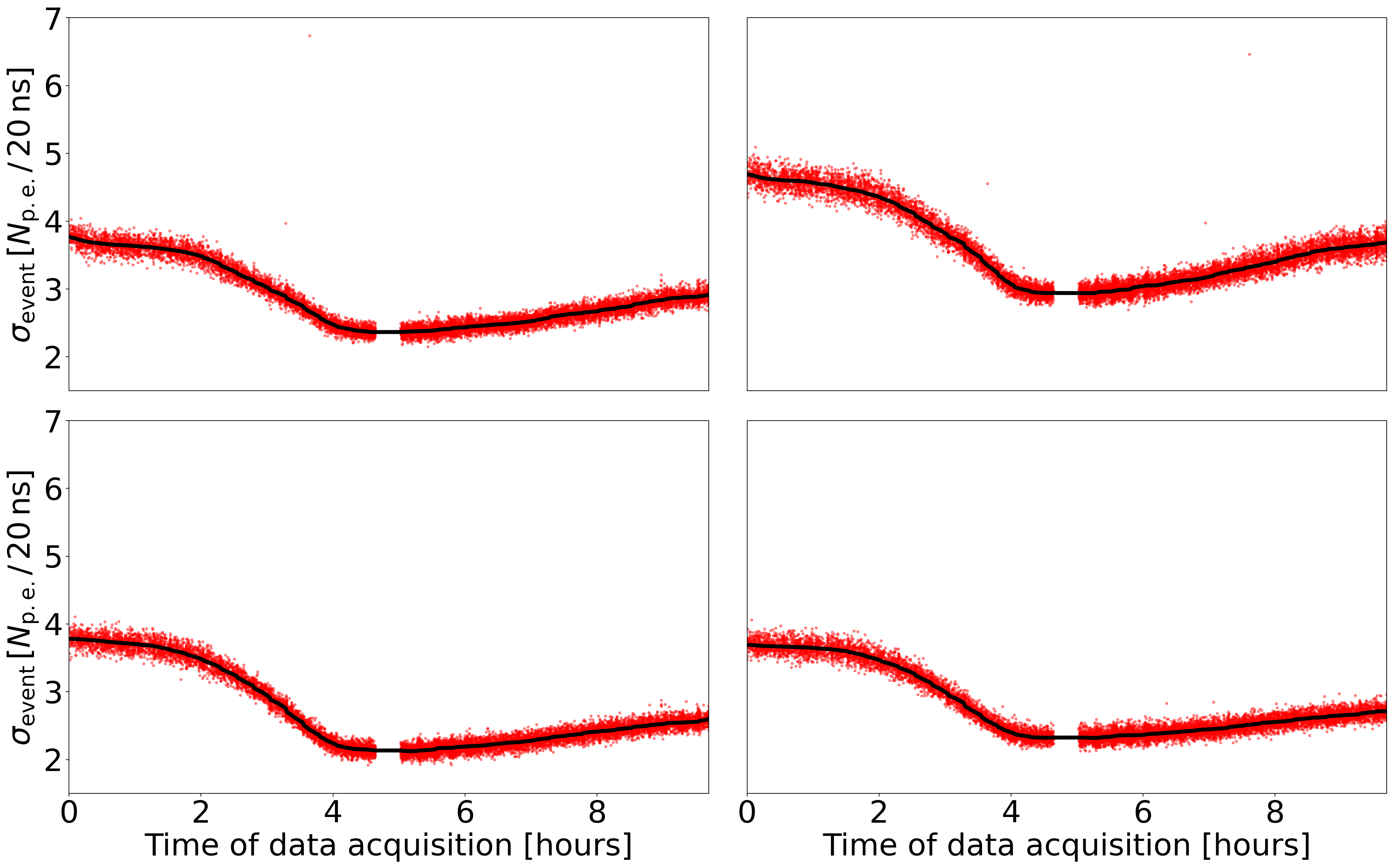}
    \caption{Standard deviations $\sigma_{\mathrm{event}}$ calculated for all $13969$ events acquired during the night of 1.9.2022. 
    Each panel corresponds to one PMT.
    The PMT layout reflects the sky view: the upper two PMTs observe elevations above $15\degree$, while the lower two observe below $15\degree$.
    The values $\sigma_{\mathrm{event}}$ are shown in red, while their smoothed average, which represents the noise background level $\sigma_{\mathrm{NB}}$, are denoted by black lines.
    }
    \label{fig:app:nb}
\end{figure}

Figure \ref{fig:app:nb} shows the $\sigma_{\mathrm{event}}$ values (colored in red) for $13969$ events collected during the night of 1.9.2022. 
Since most events do not contain significant non-noise contributions such as signal of an EAS, all are included in the analysis.
Each panel corresponds to one PMT.
It is evident that changes in the standard deviation of the NB are slow throughout the night.
This behavior is further highlighted by the black lines in each panel, which represent a smoothed estimate of the NB standard deviation~$\sigma_{\mathrm{NB}}$, derived using a combination of a moving median and a Savitzky-Golay filter \cite{Savitzky1964}.
The higher variance observed at the beginning of the night comes from background of the moonlight.
The small fluctuations around the black lines can be primarily attributed to the statistical uncertainty that comes from the estimation of the sample standard deviation.
Specifically, when the standard deviation is calculated over a finite number of samples (e.g., $1024$ bins), the result includes a non-zero estimation error given by Eq.~(\ref{eq:rel_error}).
Moreover, the presence of traces containing non-noise contributions can also contribute to these fluctuations.

This analysis has been repeated across various nights and shows consistently that changes in the NB are sufficiently slow over the night.
Its standard deviation changes gradually, justifying the simplification of the triggering algorithms described in Sect.~\ref{subsec:disc:std}.

\bibliography{main.bib}

\section*{Acknowledgements}
This work was supported by JSPS KAKENHI Grant Number 25H00647 and 21H04470.
This work was partially carried out by the joint research program of the Institute for Cosmic Ray Research (ICRR) at the University of Tokyo. 
The work at the University of Chicago was supported by NSF grant PHY-2514190. 
The Czech authors gratefully acknowledge the support of the Ministry of Education, Youth and Sports of the Czech Republic project No. CZ.02.1.01/0.0/0.0/17\_049/0008422, CZ.02.01.01/00/22\_008/0004632, LM2023032, Czech Science Foundation project GACR 23-07110S and the support of the Czech Academy of Sciences and Japan Society for the Promotion of Science within the bilateral joint research project with Osaka Metropolitan University (Mobility Plus project JSPS 21-10). 
The authors thank the Pierre Auger and Telescope Array Collaborations for providing logistic support and part of the instrumentation to perform the FAST prototype measurements and for productive discussions.

\section*{Author contributions statement}
J. Kmec performed the data analysis and wrote the manuscript.
P. Hamal performed the data analysis and Monte Carlo simulations. 
P. Hamal, M. Vacula, F. Bradfield, L. Nožka, and S. Sakurai revised the manuscript.
All authors contributed to the discussion of the results and provided comments on the manuscript.
All authors are also involved in detector construction, deployment, long-term data-taking and maintenance, software development, and manuscript review.

\section*{Data availability}
All data required to repeat the analysis and to generate all figures included in the manuscript can be downloaded from \cite{Kmec2025_data}.
Please do not hesitate to contact us if you encounter any problems while downloading the data.

\section*{Competing interests}
The contact author has declared that none of the authors has any competing interests.

\end{document}